\documentclass[sigconf, nonacm]{acmart}
\AtBeginDocument{%
  \providecommand\BibTeX{{%
    Bib\TeX}}}
\setcopyright{licensedothergov}
\copyrightyear{2025}
\usepackage[acronym]{glossaries}
\glsdisablehyper

\newacronym{NREL}{ANONYMOUS}{ANONYMOUS}

\newcommand{\marconi}{\texttt{Marconi 100}\xspace}

\newcommand{\mlscheduling}{\texttt{ML-guided scheduling}\xspace}

\newacronym{exadigit}{ExaDigiT}{ExaDigiT}
\newacronym{frontier}{Frontier}{Frontier}
\newacronym{RAPS}{RAPS}{resource allocator and power simulator}
\newacronym{S-RAPS}{\mbox{S-RAPS}}{\mbox{Scheduled-RAPS}}
\glsunset{exadigit}
\glsunset{frontier}
\newacronym{slurmsim}{Slurm Simulator}{Slurm Simulator}
\glsunset{slurmsim}
\newacronym{slurm}{Slurm}{Slurm}
\glsunset{slurm}
\newacronym{CQSim}{CQSim}{CQSim}
\glsunset{CQSim}
\newacronym{QSim}{QSim}{QSim}
\glsunset{QSim}
\newacronym{ScheduleFlow}{ScheduleFlow}{ScheduleFlow}
\glsunset{ScheduleFlow}
\newacronym{FastSim}{FastSim}{FastSim}
\glsunset{FastSim}
\newacronym{etal}{et\,al\@.}{et\,al\@.}
\glsunset{etal}
\newacronym{HPC}{HPC}{high-performance computing}
\newacronym{ML}{ML}{machine learning}
\newacronym{AI}{AI}{artificial intelligence}
\newacronym{VR}{VR}{virtual reality}
\newacronym{DT}{DT}{digital twin}
\newacronym{QoS}{QoS}{quality of service}
\newacronym{FIFO}{FIFO}{first-in, first-out}
\newacronym{DCDT}{DCDT}{data center digital twin}
\newacronym{CDU}{CDU}{cooling distribution unit}

\newacronym{ANL}{ANL}{Argonne National Laboratory}
\newacronym{BGP}{BGP}{Blue-Gene/P}
\newacronym{SWF}{SWF}{standard workload format}
\newacronym{CLI}{CLI}{command line interface}
\newacronym{TCS}{Fujitsu TCS}{Fujitsu Software Technical Computing Suite}

\newacronym{fcfs}{FCFS}{first-come, first-served}
\newacronym{sjf}{SJF}{shortest-job-first}
\newacronym{ljf}{LJF}{largest-job-first}

\newacronym{priority}{Priority}{priority based scheduling}
\newacronym{easy}{EASY}{Earliest Available Start-time Yielding}
\newacronym{ffbf}{ffbf}{first-fit backfill}

\newacronym{PUE}{PUE}{power usage effectiveness}
\newacronym{edp}{EDP}{energy-delay-product}
\newacronym{ed2p}{ED$^2$P}{energy-delay$^2$-product}

\newacronym{AWF}{$\text{AWF}$}{area-weighted average response time}
\newacronym{PWSRT}{$P^\alpha SF$}{priority weighted specific response time}
\newacronym{P2SF}{$P^2 SF$}{$P^2 SF$}
\glsunset{P2SF}

\newacronym{mlscheduling}{mlscheduling}{MLSched-KEYWORD-in abbvreviations.tex}
\glsunset{mlscheduling}
\newacronym{pm100}{pm100}{PM100}
\glsunset{pm100}

\def\code#1{\texttt{#1}}

\usepackage{flushend}
\usepackage[most]{tcolorbox}
\usepackage{amsmath}
\usepackage{xspace}
\usepackage{xcolor} 

\definecolor{NavyBlue}{RGB}{0, 102, 204}
\definecolor{Green}{RGB}{0, 153, 0}
\definecolor{Orange}{RGB}{255, 102, 0}
\definecolor{Red}{RGB}{204, 0, 0}
\definecolor{Purple}{RGB}{153, 51, 255}
\definecolor{Gray}{RGB}{128, 128, 128}
\definecolor{OliveGreen}{RGB}{128,128,0} 
\usepackage{subfigure}

\newtcolorbox[auto counter]{finding}[1][]{%
    colback=blue!5,           
    colframe=blue!40,         
    boxrule=0pt,              
    leftrule=2mm,             
    sharp corners,            
    before upper={\textbf{Finding~\thetcbcounter:}~}, 
    fontupper=\normalfont,    
}

\def\BibTeX{{\rm B\kern-.05em{\sc i\kern-.025em b}\kern-.08em
T\kern-.1667em\lower.7ex\hbox{E}\kern-.125emX}}

\newcommand{\tzi}[1]{ {\textcolor{red} {TZI-fix: #1 }}}

\usepackage{titlesec}
\usepackage[inline]{enumitem}

\usepackage{ADAE/sc25repro}

\begin{document}

\title{HPC Digital Twins for Evaluating Scheduling Policies, Incentive Structures and their Impact on Power and Cooling}

\author{Matthias Maiterth}
\authornote{Corresponding author: 
\orcid{0000-0001-8698-460X}
\textit{maiterthm@ornl.gov}}
\affiliation{%
  \institution{Oak Ridge National Laboratory}
  \city{Oak Ridge}
  \state{Tennessee}
  \country{USA}
}

\author{Wesley H. Brewer}
\affiliation{%
  \institution{Oak Ridge National Laboratory}
  \city{Oak Ridge}
  \state{Tennessee}
  \country{USA}
}


\author{Jaya S. Kuruvella}
\affiliation{%
  \institution{Texas State University}
  \city{San Marcos}
  \state{Texas}
  \country{USA}
}

\author{Arunavo Dey}
\affiliation{%
  \institution{Texas State University}
  \city{San Marcos}
  \state{Texas}
  \country{USA}
}

\author{Tanzima Z. Islam}
\affiliation{%
  \institution{Texas State University}
  \city{San Marcos}
  \state{Texas}
  \country{USA}
}



\author{Kevin Menear}
\affiliation{%
  \institution{National Renewable Energy Laboratory}
  \city{Golden}
  \state{Colorado}
  \country{USA}
}
\author{Dmitry Duplyakin}
\affiliation{%
  \institution{National Renewable Energy Laboratory}
  \city{Golden}
  \state{Colorado}
  \country{USA}
}

\author{Rashadul Kabir}
\affiliation{%
  \institution{Colorado State University}
  \city{Fort Collins}
  \state{Colorado}
  \country{USA}
}


\author{Tapasya Patki}
\affiliation{%
  \institution{Lawrence Livermore National Laboratory}
  \city{Livermore}
  \state{California}
  \country{USA}
}



\author{Terry Jones}
\affiliation{%
  \institution{Oak Ridge National Laboratory}
  \city{Oak Ridge}
  \state{Tennessee}
  \country{USA}
}
\author{Feiyi Wang}
\affiliation{%
  \institution{Oak Ridge National Laboratory}
  \city{Oak Ridge}
  \state{Tennessee}
  \country{USA}
}

\renewcommand{\shortauthors}{Maiterth et al.}


\begin{abstract}
Schedulers are critical for optimal resource utilization in high-performance computing.
Traditional methods to evaluate schedulers are limited to post-deployment analysis, or simulators, which do not model associated infrastructure.
In this work, we present the first-of-its-kind integration of scheduling and digital twins in HPC.
This enables what-if studies to understand the impact of parameter configurations and scheduling decisions on the physical assets, even before deployment, or regarching changes not easily realizable in production.
We (1) provide the first digital twin framework extended with scheduling capabilities,
(2) integrate various top-tier HPC systems given their publicly available datasets,
(3) implement extensions to integrate external scheduling simulators.
Finally, we show how to
(4) implement and evaluate incentive structures, as-well-as
(5) evaluate machine learning based scheduling, in such novel digital-twin based meta-framework to prototype scheduling.
Our work enables what-if scenarios of HPC systems to evaluate sustainability, and the impact on the simulated system. 
\end{abstract}

\ccsdesc[500]{Computer systems organization}
\ccsdesc[300]{General and reference~Cross-computing tools and techniques}
\ccsdesc[500]{Computing methodologies~Discrete-event simulation}
\ccsdesc[500]{Computing methodologies~Distributed simulation}
\ccsdesc[300]{Computing methodologies~Simulation evaluation}

\keywords{Scheduling Simulators, Digital Twin, Data Center Digital Twin, System Simulator, Distributed Systems Simulation, Batch Scheduling}


\maketitle

\section{Introduction}
The increasing complexity of highly-efficient supercomputing centers fuels an ever increasing demand for more powerful models. \Glspl{DT} have emerged as a means of integrating system telemetry, modeling and simulation, \gls{AI}, and system control mechanisms to create a virtual representation of the physical system, modelling cooling, power, and workloads \cite{athavale2024digital}.

Aiming for optimal usage of \gls{HPC} systems, different stakeholders face various challenges.
For example: \textit{users} seek feedback regarding job usage, estimated runtime, and application efficiency;
\textit{operators} monitor and tune operational parameters based on load and conditions;
\textit{center managers and vendors} seek insight into which machine aspects are the biggest barriers to performance%
, and seek trends for future procurements.
\Glspl{DCDT}
can provide estimates and simulations and even serve for design considerations and virtual prototyping of future systems~\cite{athavale2024digital}, without
consuming the system's own resources.


Scheduling is critical for efficient use of \gls{HPC}
~\cite{Potts2000,Allcock2017}.
Therefore, integrating scheduling into \glspl{DT} of \glspl{HPC} is necessary to form a representative twin of the overall system.
The sound integration of scheduling capabilities into a \gls{DCDT} extends its capability from a reactive role to a predictive one, enabling \emph{what-if} studies.
A representative yet modifiable scheduling simulator integrated into a \gls{DCDT} allows to study how a system responds to alteration of its parameters.
Production systems are not suitable for such changes, unless service interruptions are acceptable. Similarly, scheduling simulators in isolation or \gls{DCDT}'s without these capability can not answer such questions.
Integrating a scheduler into a \gls{DCDT} is therefore a valuable contribution,
revealing holistic insights into the operation of our systems and  optimization opportunities.





In this paper, we introduce \gls{S-RAPS}~\cite{S-RAPSgit}, which extends the \gls{RAPS}, originally developed for the open-source digital twin framework \gls{exadigit} by Brewer \gls{etal}~\cite{brewer2024digital}.
We present what-if studies evaluating the integration of schedulers with digital twins.
Specifically, we evaluate:
\begin{enumerate}
\item \textbf{Scheduling impact on system response} --
We explore the impact of scheduling to understand a system's power, cooling, and workload response with our integration, 
a factor not observable for scheduling simulators in isolation.

\item \textbf{Evaluation of incentive structures} -- 
We explore incentive structures and impact of imposed reward metrics on workloads and system --- a case-study impossible with the current state-of-the-art, without deploying to production.

\item \textbf{\Gls{ML} for scheduling} -- 
We study \gls{ML} guided scheduling based on metrics usually not accessible to scheduling simulators in isolation, or hardly trainable due to limited data and context without an integrated \gls{DCDT}. 

\item \textbf{Integration with external schedulers} -- 
We demonstrate the feasibility of our approach by extending scheduling-integrated \gls{DCDT} to interface with external schedulers.

\end{enumerate}

This work extends to the original work
of \cite{brewer2024digital} and use open datasets for validation and use-cases.
%
Our contributions are:
\begin{itemize}
    \item {\bf Integration of scheduling into \glspl{DCDT}}: Previous work does not integrate  scheduling simulators with digital twins. Our research is the first to enable such integration, allowing the exploration of 
    ``what-if'' scenarios to study power and cooling, given real workload and system data.
    \item {\bf Use of open datasets}: Previous work has utilized open datasets for post-mortem analysis; however, this work is the first to utilize such datasets to
    simulate the anticipated system behavior given altered scheduling parameters.
    \item {\bf Extension to other schedulers}: As each system and scheduler setup is unique,
    we provide the extensions necessary such that users can model and simulate their system and their use-cases, providing wide applicability beyond what is presented in this work.
\end{itemize}

The remainder of the paper is structured as follows:
Section~\ref{sec:Background} presents background on digital twins and scheduling,
and discusses the open data sets used in this research. 
In Section~\ref{sec:Methods}, we introduce S-RAPS. 
In Section~\ref{sec:Evaluation}, we present our evaluation and use-cases.
Section~\ref{sec:Discussion} discusses future work, and we conclude in Section~\ref{sec:Conclusion}.




\section{Background}
\label{sec:Background}
\subsection{Related Work}



\subsubsection{Digital Twins for Operational Optimization:}

\Gls{DT} research has surged in recent years, and useage of traditional modeling and simulation techniques 
have found wider adoption for operation~\cite{Tao2019DigitalTI}, using data-driven approaches~\cite{Schroeder2016} and by coupling AI with online decision making~\cite{Karkaria2025}.

In the context of \gls{HPC}, the work on \gls{exadigit}~\cite{brewer2024digital} is seminal as an open-source framework,
consisting of 
the \gls{RAPS} module, a transient thermo-fluid cooling module, and an visual analytics model of the supercomputer and central energy plant.
The work, however, only presented an initial framework with large emphasis on the thermo-fluid cooling simulator, as well as power loss modeling for job-trace replay, without venturing into the explorative aspects. 
The initial work does not include a batch scheduler and only \emph{replays} the given workloads, therefore unable to \emph{re-schedule} for what-if analysis.

In turn, our work builds upon the \gls{exadigit} framework, with explicit focus on scheduling, introducing \gls{S-RAPS}.
We extend \gls{exadigit} to support open data sets for additional systems, and explore use-case driven analysis for \gls{HPC}, and evaluate the impact of various scheduling policies on power and cooling.

\subsubsection{Scheduling Simulators:}
\label{BG:SchedSim}
Simulating scheduler behavior has been an active area of research, consistently
supporting advances in \gls{HPC}~\cite{Boezennec2024}.
Popular examples of batch scheduling simulators are
\gls{slurmsim}~\cite{Simakov2018SlurmSim, Simakov2018-2,Simakov2022},
and scheduler specific alternatives such as the work by Wilkinson \gls{etal}~\cite{Wilkinson2023FastSim}.
\gls{CQSim}~\cite{CQSim}, which originated in \gls{QSim} is a prominent example outside of the \gls{slurm} ecosystem.
%
This is a non-exhaustive list, as simulators such as
GridSim~\cite{Buyya2002gridsim},
SimGrid~\cite{Casanova2014SimGrid},
Bricks~\cite{Takefusa1999Bricks},
Simbatch~\cite{gay2006simbatch},
Alea~\cite{Klusavcek2019alea},
AccaSim~\cite{Galleguillos2018},
BBSched~\cite{Fan2019BBSched},
ScSF~\cite{Rodrigo2017scsf},
Batsim~\cite{Dutot2016},
as well as schedulers that have built in simulators, such as
Torque/Maui~\cite{Jackson2001Maui} and Moab Scheduler, played an important role in the development of scheduling simulators and should not be left unmentioned.



These simulators generally are not focused on the systems infrastructure but the core of scheduling.
With continuous progress in scheduling simulators in general, the aim is to enable the integration of such advanced scheduler developments into \Glspl{DT}. 
We designed \gls{S-RAPS} to leverage existing work such that users can interface with other scheduling simulators, such as the 
\gls{slurmsim} or \gls{FastSim}.
This enables easy extensions to S-RAPS for studying power and cooling.
We demonstrate an integration of \gls{FastSim}~\cite{Wilkinson2023FastSim} into \gls{S-RAPS} in Sect.~\ref{sec:Evaluation:Schedulers:FastSim}. 

\subsection{Open Datasets}
\label{sec:BG-Datasets}
\label{sec:Background:Datasets}

\begin{table*}[th]
    \centering
    \caption{Systems and datasets used in study.}
        \begin{tabular}{lllllll}
        \toprule
        \textbf{System} & \textbf{Architecture} & \textbf{Nodes} & \textbf{Dataset}& \textbf{Scheduler} & \textbf{Job Count} & \textbf{Characteristics} \\
        \midrule
        Frontier & HPE/Cray EX & 9600 & Proprietary & Slurm & 1,238 & job traces (15s), CPU/GPU power \& temp. \\
        Marconi100 & IBM POWER9 & 980 & PM100 \cite{antici2023pm100} & Slurm & 231,238 & job traces (20s), CPU/node power \\ 
        \hline
        Fugaku & Fujitsu AF64FX & 158,976 & F-Data \cite{antici2024fdata} & Fujitsu TCS & 116,977 & job summary, node-level power only \\
        Lassen & IBM POWER9 & 792 & LAST \cite{patki2021monitoring, last2024} & LSF & 1,467,746  & job summary, includes network tx/rx \\
        Adastra & HPE/Cray EX & 356 & Cirou \cite{Adastra15D} & Slurm & 30,570 & job summary,  job avg component power \\
        \bottomrule
    \end{tabular}

    \label{tab:datasets}
\end{table*}


In this work, we selected four open datasets for accessibility and reproducibility, as shown in Table~\ref{tab:datasets}: 
PM100 \cite{antici2023pm100} from Marconi100, 
F-Data \cite{antici2024fdata} from Fugaku, 
LAST \cite{last2024} from Lassen, and
Cirou's dataset \cite{Adastra15D} from Adastra.
%
Additionally, we use a proprietary dataset from Frontier due to its extensive verification and validation in the context of \glspl{DCDT} from previous work
~\cite{brewer2024digital} for reproducibility.

The telemetry traces used for simulation vary according to the data source, which we discuss as follows:
\noindent$\bullet$~\textbf{Marconi100:}
The Marconi100 system at CINECA has two public datasets: the M100~\cite{borghesi2023m100} and the PM100 dataset~\cite{antici2023pm100}.
We use the PM100 data 
as it is pre-curated.
We filter jobs containing shared nodes as this is not yet supported in our model.
The data includes CPU, memory and node power in a 20-second interval.
As the data has been filtered, it does not reflect the system's full operational utilization.
This means that replay and reschedule will differ~\cite{antici2023pm100}.

\noindent$\bullet$~\textbf{Fugaku F-Data}:
F-Data~\cite{antici2024fdata} is a dataset containing job and performance information with derived metrics for job classification from Fugaku.
It includes monthly data from March 2021 to April 2024, with the following 
job metrics: 
energy consumed, node power (minimum, maximum and average),  
performance characteristics on operations, memory activity and the resulting performance class identified as either \textit{compute-} or \textit{memory-bound}.
%

\noindent$\bullet$~\textbf{Lassen LAST}:
This is a 1.4-million-job dataset from the Lassen supercomputer~\cite{last2024}.
It includes information on job allocation, node allocation, and job-step disposition. These are combined to get usable information for each job allocated with accumulated energy data. 
Lassen uses the IBM's LSF Job scheduler~\cite{patki2019comparing}.

\noindent$\bullet$~\textbf{Adastra:}
CINES has published 15 days of the Adastra system~\cite{Adastra15D}, 
including node power, memory power, and CPU power.
The heterogeneous system has a CPU and GPU partitions.
GPU power is not provided, but can be derived from node power and the other components.
It uses Slurm \cite{alaoui2023porting}, 
but no scheduling policy is stated.

\noindent$\bullet$~\textbf{Frontier:}
The Frontier dataset was initially used to develop the \gls{exadigit} \gls{DCDT}.
The dataset is an excerpt from the center's continuous collection, obtained from both Slurm data, as well as Cray EX Telemetry API, collected in the STREAM system~\cite{Adamson2023}.
The scheduling policy is a priority-based mechanism that uses a modified \gls{FIFO} queue, boosted based on node count and penalized on allocation overuse~\cite{FrontierUserGuide}.
This dataset is the only one not publicly available. 
Since it was used for initial verification of \gls{RAPS}\cite{brewer2024digital}, it was important for cross validation of \gls{S-RAPS}.

\section{Method \& Design}
\label{sec:Methods}

In the following, we show the current state-of-the-art, the \gls{exadigit} framework~\cite{brewer2024digital}, with its \acrfull{RAPS}, for context, and present our  \acrfull{S-RAPS} extension. 
We begin by discussing the mechanisms of the existing simulation loop of the forward-time \gls{DCDT} simulator by Brewer \gls{etal}~\cite{brewer2024digital}.
We then present \gls{S-RAPS}, with its built-in scheduler and how this ties into the \gls{DCDT} simulators. We then show how \gls{S-RAPS} extends to external forward-time or event-based simulators.
Finally, we show how extensions for dataloaders allow us to load and simulate diverse datasets and systems, showing the true value of an open-source digital twin framework.

\subsection{Prior state of ExaDigiT}
\label{sec:Methods:PriorArt}
The original design of \gls{exadigit} consists of three main modules~\cite{brewer2024digital}: 
\begin{itemize}
    \item[(1)] Modelica-based cooling model
    \item[(2)] \Acrfull{RAPS}
    \item[(3)] Visual analytics model 
\end{itemize}
\begin{figure}
    \centering
    \includegraphics[width=\linewidth]{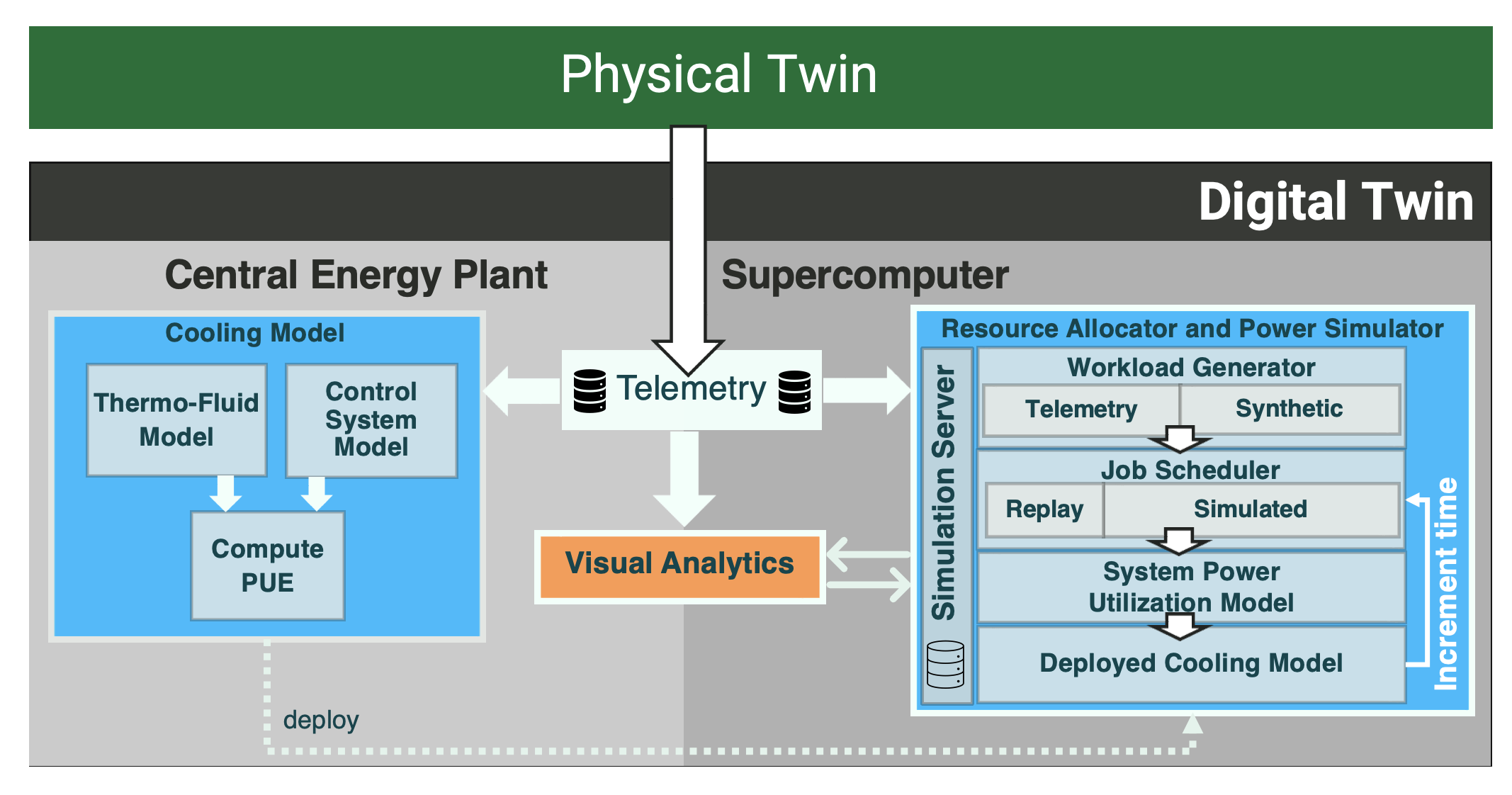}
    \caption{Simplified Original\gls{exadigit} overview in accordance with Brewer \gls{etal}~\cite{brewer2024digital}, with \gls{RAPS} module on the right.}
    \label{fig:arch2024}
\end{figure}
\Gls{exadigit}'s simulation
is depicted in Fig.~\ref{fig:arch2024}.
The digital twin reads a sequence of job traces or telemetry and 
placed on the system simulator as recorded.
For each timestamp, the observed node utilization is replayed. 
The simulated utilization is converted to a power profile, with power rectification and conversion losses applied~\cite{Wojda2024}. 
The power and therefore generated heat is fed into a cooling model~\cite{Kumar2024,Greenwood2024}.
The cooling model simulates from ~\gls{CDU} to cooling towers, giving an accurate representation of the system at each timestep. 

In the simulation, the \gls{RAPS} module provides the inputs to the cooling model and is also the main driver of the simulation loop.
This is outlined in Algorithm 1 of~\cite{brewer2024digital}:
\begin{enumerate}
    \item Initialization of system and data, and start of simulation.
    \item Simulation loop:
        \begin{enumerate}
            \item Addition of newly arriving jobs to the job queue
            \item ScheduleJobs: selection and placement of available jobs to available resources.
            \item Tick: management of resources, and calculation of compute resource utilization, power and cooling
        \end{enumerate}
\end{enumerate}
The original work processes jobs in a scheduler class, which contains the \gls{DT}'s replay mechanism.
As shown in Fig.~\ref{fig:arch2024}, this only considers replay of recorded \textit{telemetry} or \textit{synthetic} data, not scheduling.
For a generic \gls{HPC} \glsentrylong{DT}, the ability to alter the scheduling \emph{policy} and resulting job placement is key.
For this, we describe the refactoring necessary to enable generic built-in scheduling and allow for the integration of external schedulers with \gls{S-RAPS}, enabling the scheduling scenarios we present in this paper. 

\subsection{S-RAPS: Scheduled - Resource Allocator and Power Simulator}
\label{sec:Methods:Improvement}
We now present the improved simulation loop for comprehensive integration of schedulers within \gls{exadigit}'s \gls{RAPS}, named \gls{S-RAPS}.
\begin{figure}
    \centering
    \includegraphics[width=\linewidth]{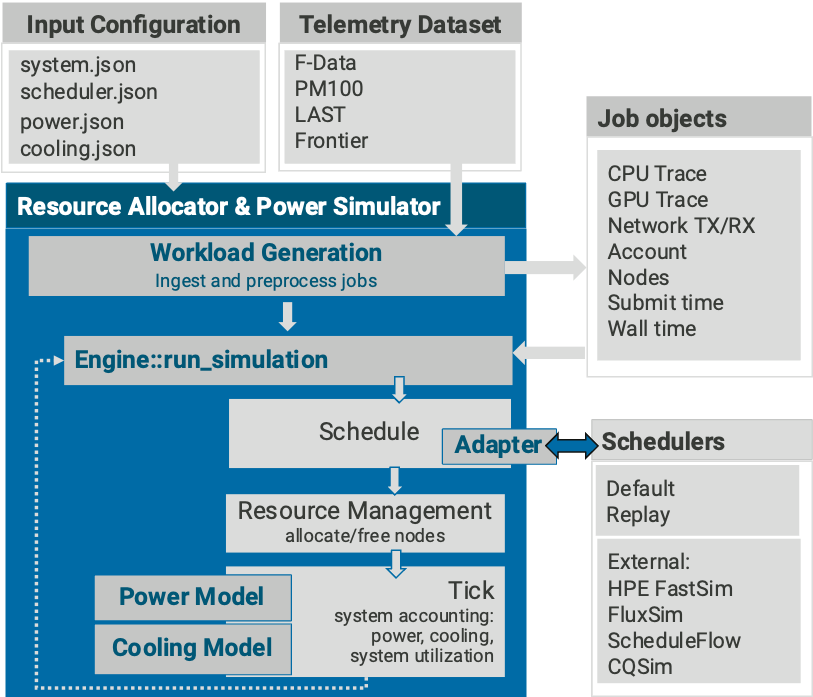}
    \caption{\acrfull{S-RAPS}: Integration of scheduling into the design of  \gls{exadigit}'s \acrfull{RAPS}. With improved configuration mechanisms, pluggable dataloaders, interface to build-in and externals schedulers, and overhauled simulation loop.}
    \label{fig:arch}
\end{figure}
Figure~\ref{fig:arch} shows the overhauled design of \gls{S-RAPS}, enabling the integration of forward-time or event-based schedulers.
Key changes include refactoring and generalizing of the following five components:
\begin{enumerate*}
    \item system initialization,
    \item dataloaders,
    \item simulation engine,
    \item scheduler abstraction,
    \item systems accounting
\end{enumerate*}.
This is accomplished while keeping the original \gls{RAPS} simulation concept with intended scheduling, resource management, and \texttt{tick} intact.

\subsubsection{System initialization:}
\label{sec:Methods:Improvement:Initialization}

The system initialization has been improved to capture not only the system's configuration, but also to include information necessary for scheduler simulation.
The rework introduces cleaner abstractions and separation of concerns, with future extensions in mind. 
The core loop of the simulator was refactored introducing a simulation engine.
During system initialization, telemetry is used to initialize the job objects augmented with information for scheduling.
The telemetry is now also used to initialize user-account information (clear or anonymized).
The refactored simulation engine separates resource manager and scheduler interface, which loads either the build-in or external scheduler.
Finally, objects for tracking statistics of the simulation are initialized.

The \gls{HPC} System configurations, their dataloaders, and the schedulers are implemented as plugins.
The specific configurations are selectable on simulation start via the \gls{CLI} and are designed to be easily extensible.
This helps with  
rapid testing of configuration and experimental design: administrators can easily represent their systems, and developers of scheduling simulators can easily load their policies, once extended.


\subsubsection{Dataloaders:}
\label{sec:Methods:Improvement:Dataloaders}
A dataloader's task is to load and parse the telemetry data and generate the list of to-be-scheduled jobs.
Each job requires information on: \textit{submit time}, \textit{start time}, \textit{end time}, \textit{time limit}, and the \textit{number of requested nodes} (alternatively, the exact set of nodes to which the job was assigned).
This is a standard for scheduling simulators as for example used in the \gls{SWF}~\cite{Chapin1999}. 
The dataloaders also load the job traces for replay in the \gls{DCDT} simulation. 
If traces are not available for a dataset, scalar values can for example represent a job's average power, energy, or other characteristics -- depending on data availability.

When rescheduling, the job traces recorded may not coincide with the time slice needed for the simulation.
We treat such occurrence as missing data, using the last known value. 
Therefore, for correct simulation the dataloader must identify the following key times for each job:
\begin{itemize}
    \item Job submit, start and end time
    \item Telemetry start and end time
\end{itemize}
Given the individual job times and the overall timespan of the dataset, we can run the simulation within the range of the overall telemetry.
Additionally, this change enables us to either replay the recorded data as is or simulate a new schedule.

\begin{figure}[t]
\includegraphics[width=\columnwidth]{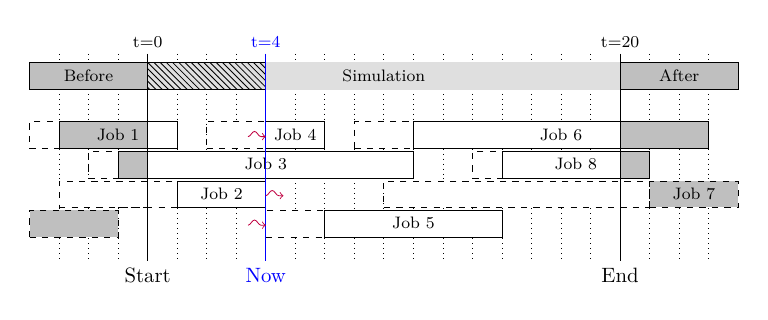}
    \caption{Example job trace, with job-submit time, -start time and -end time.
    The time-stepped simulator triggers on each time step, while the event based scheduling simulator only has to react to triggered events (magenta arrows) such as start of a job (job 4), end of a job (job 2), and submission of a new job (job 5).}
    \label{fig:sim}
\end{figure}

An example of such timeline is presented in Fig.~\ref{fig:sim}, showing a simulation from $t=0$ to $t=20$, with submission times of jobs indicated by dashed outline extending in front of the jobs, and actual execution indicated by the solid outline of the jobs. During rescheduling, the jobs can be placed as early as they have been submitted, which is ultimately decided by the selected scheduling policy. Jobs that ended before start of the simulation time or were submitted after end of the simulation time are dismissed.
%
The simulation of the power and cooling behavior can then be simulated as implemented in \texttt{tick} of \gls{S-RAPS}, with the modified timeline.\footnote{There are two edge cases to consider:
jobs which originally started before the capture time (see Fig.~\ref{fig:sim}, Job 1) and jobs which ended after the capture time (see Fig.~\ref{fig:sim}, Jobs 6, 7, 8). 
When simulating a new schedule, these jobs may therefore have no corresponding telemetry at the associated simulation times.
Therefore, when rescheduling these jobs within the simulation time, these cases need to be flagged as they can cause potential discrepancies in other simulations as no known ground truth is available to \gls{S-RAPS}.} 

\subsubsection{Simulation engine:}
The simulation engine contains the main simulation loop.
At start of the simulation, the user selects explicit simulation start and end time.
Given this information, the system state is prepared, and
in case the dataset contains jobs before the selected simulation start, these jobs are placed to prepopulate the system.
This allows us to represent the actual system condition as observed in the telemetry at start of the simulation%
%
\footnote{This is often neglected by scheduling simulators,
        which ignore jobs before simulation start,
        and therefore need time to fill up the queue and system, distorting the results.}$^{,}$\footnote{This also allows \gls{S-RAPS} to simulate anticipated schedules and profiles from live-data.}.
We can then enter the main simulation loop.
It is refactored into four well-defined steps: 
        \begin{enumerate}
            \item Preparation of the time step:
                Before each iteration, the system state is updated.
                For this, completed jobs are cleared from the system, freeing resources, and updating the state of nodes.
            \item Addition of eligible jobs to the job queue: 
                Jobs that have been submitted --- according to simulation time --- are added to the job queue.
                %
                %
            \item Call of \texttt{schedule}: 
                The jobs are scheduled according to the job queue and selected scheduling policy,
                and placed on the system in coordination with the resource manager.
                %
            \item Call of \texttt{tick}:
                The engine's \texttt{tick} function calls the sequence of \gls{DCDT} simulators and models, and increments the time. 
        \end{enumerate}
These descriptions seem simple, but required major changes from the simple replay mechanism of the original design.
For example, in the original replay mechanism, node placement was not enforced. 
In the overhauled design, the exact node placement as specified in the telemetry is used in replay mode.
However, when rescheduling the scheduler select the appropriate nodes. 
The resource manager then completes the job placement, allocating nodes.
Additionally, the refactor resolved timing and allocation issues for nodes with both ending and starting jobs coinciding in the same time step. 

Regarding job eligibility, in the original replay mode, all jobs were part of the job queue. Jobs were then placed on the system as soon as indicated by their start time.
With the updated design, jobs can only be scheduled and placed once they have been submitted.
This also means that pre-computing a schedule is not possible as the digital twin observes the jobs as they are submitted, just like a real system.
The scheduler is not aware of jobs not yet in the queue.

When calling the \texttt{schedule} function the loaded implementation is triggered.
This recomputes the order of the job queue according to selected policy and coordinates with the resource manager to place eligible jobs.
The split of resource manager and scheduler via this interface was a major improvement, separating the built-in scheduling capabilities and enabling the use of external schedulers%
\footnote{The original design included a reschedule functionality, however it simply redistributed the job start times according to a Weibull distribution and was not representative of batch scheduling.}.



The update to \texttt{tick}
now also represents a clear separation of concerns
enabling easier replacement of simulation sub-modules, where tick is only responsible for the simulation of physical sub-systems.
Our redesign of \gls{S-RAPS} with the simulation engine puts strong focus on extensibility and the use of plug-ins. This also enables support for future site-specific customizations.


\subsubsection{Scheduler abstraction:}
\label{sec:Methods:Scheduler-Abstraction}
As outlined, the simulation engine triggers the scheduler in each iteration of the simulation loop.
Any external scheduler and scheduling simulator has its own set of logics regarding which events to track and react to.
The abstraction we use enables users who interface \gls{S-RAPS} with their scheduling simulator to implement the logic for triggering and sending these events.
 Figure~\ref{fig:sim} illustrates such case, with the triggered evens for time step $t=4$ shown as magenta arrows.

\Gls{S-RAPS} interfaces with the scheduler in case the simulator provides new information in a given iteration:
(1) by triggering the scheduler to recompute the schedule, or deciding to skip if  no change has occurred;
(2) it interprets the information returned from the scheduler; 
and (3) \gls{S-RAPS} then triggers the resource manager, placing identified jobs on the system, and maintains the job queue.
This ensures that the remainder of the simulation can progress.

\subsubsection{Built-in and external schedulers:}

The scheduler can be selected via the \code{\lstinline{--scheduler}} \gls{CLI} option.
Its policies are selected via the \code{\lstinline{--policy}} and \code{\lstinline{--backfill}} options.
The default is the build-in scheduler which implements the policies:
\gls{fcfs},
\gls{sjf}, 
\gls{ljf}, and
priority-based scheduling.
Additionally, the default scheduler also provides the replay mechanism of the original \gls{RAPS} implementation.
Regarding backill options the supported defaults are 
\textit{no-backfill}, 
\textit{first-fit}, and 
\textit{easy} (i.e. \gls{easy}\cite{EASY}).
All options are extensible for use with external schedulers.

While the default scheduler provides basic scheduling policies,
it does not provide implementations for best-fit, greedy, conservative, or other more sophisticated implementations.
For this, we provide the interface for external schedulers and scheduling simulators.
As shown in Sect.~\ref{BG:SchedSim}, the numerous schedulers and scheduling simulators all have their validity, and we do not compete but try to enable them, and provide example integration in the source-code.

\subsubsection{Systems accounting:}
The final major rework to discuss is the addition of system accounting and statistics.
The original \gls{RAPS} design kept track of general simulation and HPC system
statistics, with focus on the power and cooling simulation of the \gls{DCDT}.
\gls{S-RAPS} extends those and adds collection of statistics for jobs, users, accounts, as well as scheduler-focused statistics. 
%
This allows users to easily extend metrics of interest for their facility or experiments.

Previously tracked information includes:
completed jobs,
job throughput,
average system power, 
power loss, 
system power efficiency,
total energy consumed, 
and the cost estimates for carbon emissions. 
\gls{S-RAPS} adds more scheduler-specific information and also aggregates according to user accounts, 
such as (non-exhaustive):
queued and running jobs 
average job size, 
histogram of job size scheduled (small, medium, large, by node count),
aggregate node hours,
average power and energy per job, their
\gls{edp}, \gls{ed2p},
average CPU and GPU utilization,
wait time, 
turnaround time, 
as well as area weighted response time (the average turnaround time per unit of node-hour across all scheduled jobs), and priority-weighted specific-response time (average sensitivity-adjusted turnaround time per unit of node-hour), which helps to capture packing efficiency and fairness~\cite{goponenko2022metrics}.

By tracking this information for both the system and user accounts, we can assess if a setting of the scheduler favors specific jobs or users.
The generated statistics can be compared and correlated within a single simulation and across multiple simulations.
This allows us to investigate, e.g. how changes of the job-mix are related to job-turnaround time and observed power swings.
%
For the user account metrics, we added the option to store and reload collected user account statistics at the start of a run, supporting aggregation of this information across simulations.


In summary, these changes establish capabilities for extracting broader and deeper insight about jobs, users, and the system, which standalone scheduling simulators cannot provide without integration into a \gls{DCDT} framework. 
It is worth emphasizing that the holistic modeling of power, cooling, and job behavior relies on the integrated design of the full DCDT (Fig.~\ref{fig:arch2024}), where interactions between subsystems are critical. Such cross-disciplinary dynamics cannot be replicated by aggregating telemetry data in isolation, even with comprehensive scheduling records.

\section{Evaluation and Use-Cases}
\label{sec:Evaluation}
In the following,
we evaluate the S-RAPS scheduler interfaces and extensions, utilizing datasets from a diverse set of \gls{HPC} systems and external schedulers. 
In Sec.~\ref{sec:Evaluation:Dataloaders} we evaluate the rescheduling mechanism with 
different policies and datasets.
This is followed by use-cases:
interfacing and integration with external schedulers, such as ScheduleFlow and FastSim in Sec.~\ref{sec:Evaluation:extSchedulers};
incentive structures in Section~\ref{sec:Evaluation:Incentive}; 
and evaluating of ML-guided scheduling in Sec.~\ref{sec:Evaluation:ML}.


\subsection{Evaluation of the built-in scheduler}
\label{sec:Evaluation:Dataloaders}

\begin{figure}
    \includegraphics[width=\columnwidth]{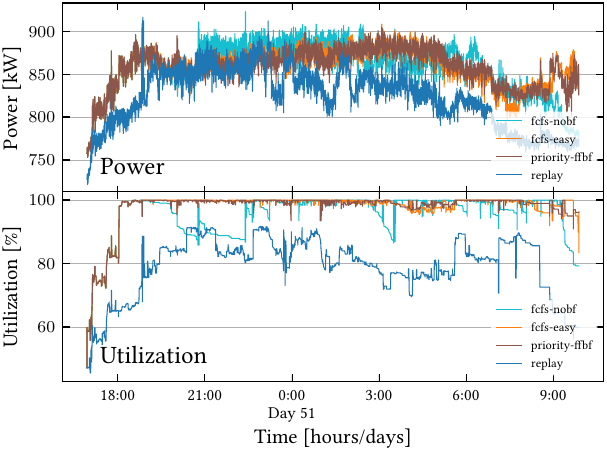}
    \caption{Replay and reschedule of the data from the PM100 Dataset (offset 50 days +17h). Showing \gls{fcfs} with no backfill (fcfs-nobf), \gls{fcfs} with \gls{easy} backfill (fcfs-easy), priority scheduling with first-fit backfill (priority-ffbf) and replay as jobs were executed, for system power and utilization.}
    \label{fig:Marconi100-day50}
\end{figure}

\begin{figure}
    \includegraphics[width=\columnwidth]{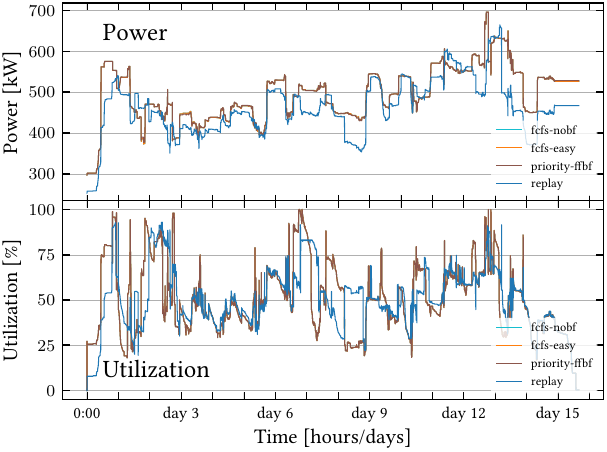}
    \caption{Replay and Reschedule of 15 days of Adastra (full dataset \cite{Adastra15D}). Replay is shown in blue, while all rescheduled runs (\gls{fcfs} \& priority) overlap almost exactly (brown line).  Given known job-power profiles and schedule information, the simulator can predict and match the observed power profile, seen as matching timed up/down-swings.}
    \label{fig:Adastra15days}
\end{figure}

\begin{figure}
    \includegraphics[width=\columnwidth]{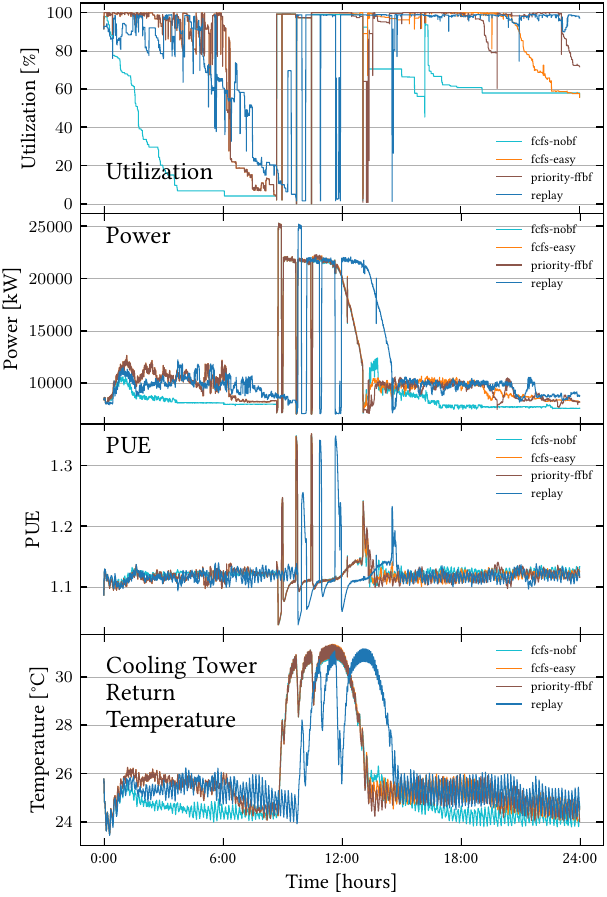}
    \caption{Replay and reschedule of the data used in \cite{brewer2024digital}.
    Showing \gls{fcfs} with no backfill (fcfs-nobf), \gls{fcfs} with \gls{easy} backfill (fcfs-easy), priority scheduling with first-fit backfill (priority-ffbf) and replay.
    The plots show system utilization, system power, as well as \gls{PUE}\textsuperscript{8} and cooling tower return temperature as simulated given the cooling model provided by~\cite{Kumar2024}.}
    \label{fig:Frontier0118}
\end{figure}

We implemented dataloaders for Frontier, Marconi100, Lassen, Fugaku, Adastra, given the datasets of Table~\ref{tab:datasets}.
Each system provides slighly different information with telemetry for Adastra, Fugaku, and Lassen providing average values for utilization, while Marconi100 and Frontier provide traces for their jobs' resource utilization.
A system and its associated dataloader is selected with the \code{\lstinline{--system}} \gls{CLI} option.
We present Figs. \ref{fig:Marconi100-day50}, \ref{fig:Adastra15days}, and \ref{fig:Frontier0118}, where
each plot shows full-system power as calculated by the power model and system utilization according to node occupancy%
\footnote{The datasets do not contain reservations and job dependencies, nor was information about down or drained nodes available. This information could greatly increase the accuracy of schedules, especially when linking the \gls{DCDT} to a live system.}.
Figure~\ref{fig:Frontier0118} additionally shows \gls{PUE} and the water temperature arriving at the cooling towers, as simulated by the cooling model. 
We show 
    replay according to the telemetry (\textit{replay}, blue),
    as well as \gls{fcfs} scheduling (\textit{fcfs-nobf}, teal),
    \gls{fcfs} scheduling with \gls{easy} backfill (\textit{fcfs-easy}, orange),
    and priority scheduling with \gls{ffbf} (\textit{priority-ffbf}, brown).
Priorities are used as provided by the datasets and respective documentation.


Figure~\ref{fig:Marconi100-day50} shows day 50 of the PM100 dataset, from 17:00 to the next day at 10:00.
The \textit{replay} utilization curve (blue) is near 80\%, with a filling job queue.
The rescheduled runs achieve very high utilization at 100\% continued utilization using backfill.
In the plot, the system shows higher aggregate power in the non-backfilled approach (teal).
The statistics show that average power consumption ($-2\%$) per job and job size ($-5\%$) decreased using either backfilled policy. 
In combination, the adjusted job placement and start times result in smoothing of the aggregate load, mitigating the power jump at 21:00, observed in the non-backfilled schedule. 

Figure~\ref{fig:Adastra15days} shows 15 days of replay and reschedule of the Adastra dataset. As the system utilization is lower and queues not filling up, the choice of scheduling algorithm makes little difference. Noteworthy is that with information on the jobs' power profiles and correct estimates of runtimes, the \gls{S-RAPS} simulator can match the observed changes in both utilization and simulated power. 

Figure~\ref{fig:Frontier0118} shows the same snapshot, as presented in the original paper by Brewer \gls{etal}~\cite{brewer2024digital}, with the cooling model of  Kumar \gls{etal}~\cite{Kumar2024}.
We apply the same scheduling policies as used in the previous cases.
The utilization plot shows that the system is making space for three full-system runs, emptying the nodes. Then, the three full-system runs are executed, and afterwards a normal job mix of varied size and lower total power is observed. 
The different power, \gls{PUE}%
\footnote{Power is modeled using~\cite{Kumar2024}. PUE for the actual system is at an average of 1.06\%.}
and return temperature behavior for the different scheduling policies is clearly visible with regard to how each policy clears the system for the large scale runs.
Regarding differences of replay to reschedule, \gls{S-RAPS} is able to place the large 9216 node jobs earlier (\textit{fcfs-nobf}, \textit{fcfs-easy} and \textit{priority-ffbf}, all overlap and start them at the same time).
While freeing nodes for the large runs, the backfilled policies are able to achieve higher utilization compared to \textit{replay}.
This, however, is due to the fact that we do not have access to node status, such as information on down or drained nodes.
The backfilled policies smooth out the power (and cooling temperature) jump observed in the \textit{fcfs-nobf} case, after the large runs, in similar fashion as described for Fig.~\ref{fig:Marconi100-day50}.

\subsection{Evaluation with external schedulers}
\label{sec:Evaluation:extSchedulers}
The integration of external scheduling simulators --- ScheduleFlow and FastSim --- serve to highlight opportunities for community extensions and future exploration.
%

\subsubsection{ScheduleFlow Scheduler:}
\label{sec:Evaluation:Schedulers:ScheduleFlow}
To prototype the integration of external schedulers, we implemented an interface to \gls{ScheduleFlow} by Gainaru \gls{etal}~\cite{Gainaru2019}.
The scheduler is event-based and maintains its own internal system state.
We implement the interface as described in Sec.~\ref{sec:Methods:Scheduler-Abstraction} and trigger the necessary internal \gls{ScheduleFlow} functionality. 
Hereby, we couple the event-based scheduler of \gls{ScheduleFlow} with the forward-time simulation of \gls{S-RAPS}.
As \gls{ScheduleFlow} is not designed for this use-case, nor optimized for performance, this initiates frequent recalculation of the schedule incurring large overheads.
The proof of concept was evaluated using synthetic runs, but shows poor performance for any of the real datasets.
The main purpose is however achieved: we are able to trigger external schedulers, which allows them to interact with the \gls{DCDT} simulations made available via \gls{S-RAPS}.
This serves as template for other schedulers, as successfully demonstrated by FastSim.


\subsubsection{FastSim Scheduler:}
\label{sec:Evaluation:Schedulers:FastSim}
\begin{figure}
    \includegraphics[width=\columnwidth]{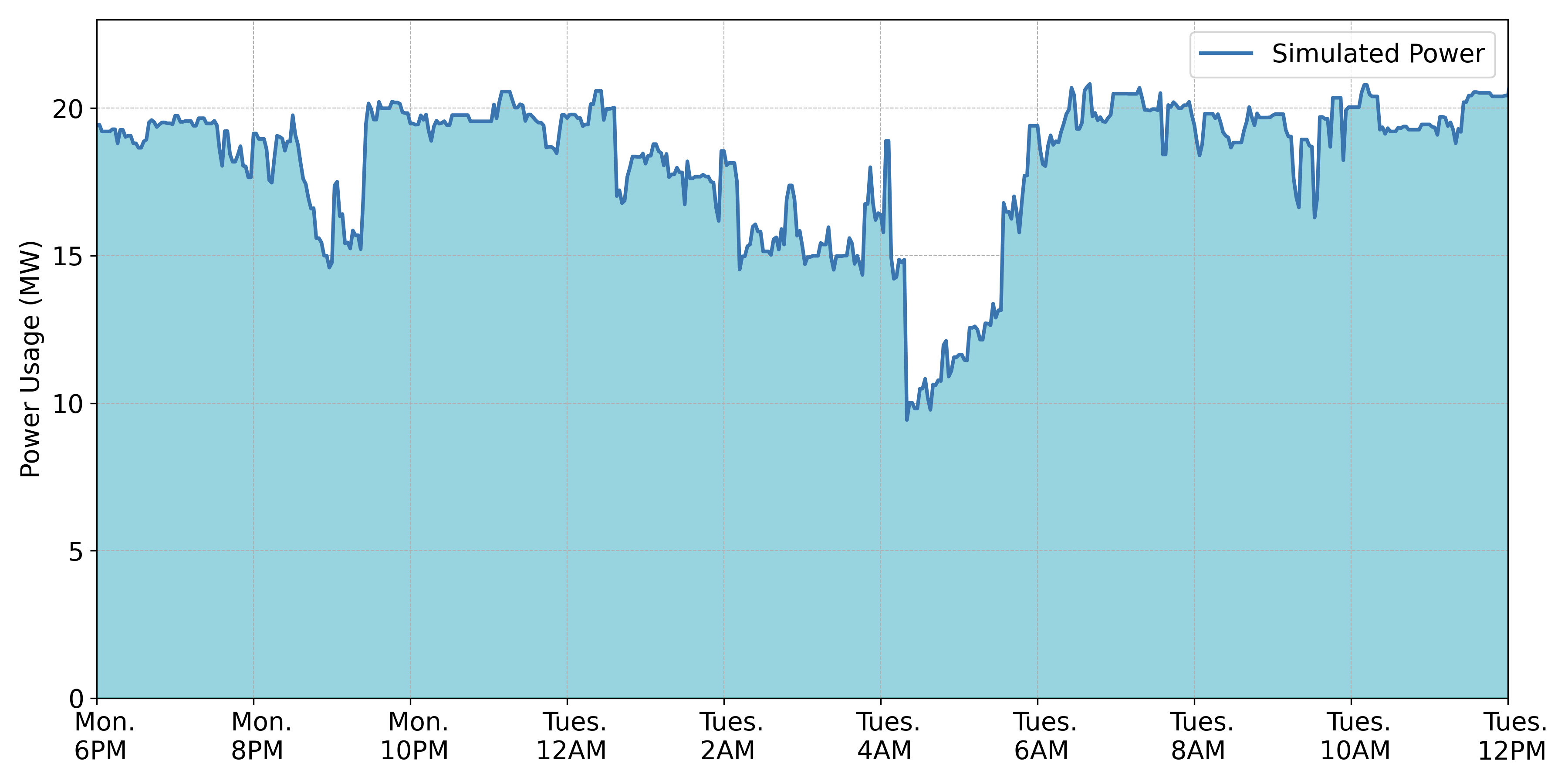}
    \caption{The results of simulating a synthetic job trace running on Frontier with the FastSim scheduler. The simulated job schedule is passed to ExaDigiT, which can then compute the resource usage over time.}
    \label{fig:fastsim_power}
    \vspace{-0.2in}
\end{figure}

FastSim~\cite{Wilkinson2023FastSim} is a lightweight emulation 
    of the Slurm scheduler software.
This external tool can simulate 
    cluster behavior up to
    thousands of times faster than real-time.
This integration moves the ExaDigiT \gls{DCDT}
    towards the capacity to
    forecast future events.
In Fig.~\ref{fig:fastsim_power}, we show
    a dip followed by a spike in Frontier's power usage
    on Tuesday morning,
    as simulated by FastSim 
    using a synthetic job trace
    developed for Frontier based on
    the workload statistics in~\cite{brewer2024digital}.
Accurate forecasting of such events 
    can inform energy-aware scheduling
    to mitigate the effects of such
    significant fluctuation in the power draw.

To integrate this simulator with \gls{S-RAPS},
    FastSim was modified by
    developing a plugin mode option.
When operating in this plugin mode,
    FastSim responds to
    a request for the system state at 
    a time step specified by 
    the driving simulator. 
FastSim then processes any events 
    which have occurred 
    up until the requested time step 
    and responds with 
    a list of running jobs 
    indexed by job ID.
When triggered by \gls{S-RAPS}, the returned list is used 
    to allocate resources and subsequently continue
    in the simulation procedure of \gls{S-RAPS}.
This process requires both 
    \gls{S-RAPS} and FastSim to maintain 
    separate copies of the system state,
    which reduces communication between 
    the two simulators 
    at the cost of
    additional computational overhead.
While this process is effective for the real-time
    simulation necessary for a \gls{DT},
    for the purpose of historical job trace rescheduling,
    we found it was faster to run FastSim and RAPS sequentially,
    with FastSim handling the job scheduling and RAPS managing the resources.
To generate the results seen in Figure~\ref{fig:fastsim_power},     a synthetic job trace of 5,324 jobs run over a period of 
    15 days was simulated using this sequential approach.
The entire simulation time was completed in 31 minutes and 24 seconds,
    amounting to a simulation speedup of 688x compared to real time.

\subsection{Use-Case: Incentive Structures}
\label{sec:Evaluation:Incentive}
By incorporating information on user accounts, \gls{S-RAPS} is able to mimic implementations of scheduling incentives and their impact on jobs and system.
In the presented use-case, we study the ability to calculate a Fugaku point score according to Sol\'orzano \gls{etal}~\cite{Solorzano2024}.
For any completed job, its statistics are accumulated to the issuing account.
This information can be accumulated across multiple simulations or used for prioritization, as described in the redeeming phase of~\cite{Solorzano2024}.
The implementation is added in \textit{schedulers/experimental.py} of \cite{S-RAPSgit}, which has policies to derive priorities from account based on: \textit{Fugaku Points}, \textit{power} usage, accumulated \gls{edp}, \gls{ed2p}, and others. 

\begin{figure}
    \includegraphics[width=\columnwidth]{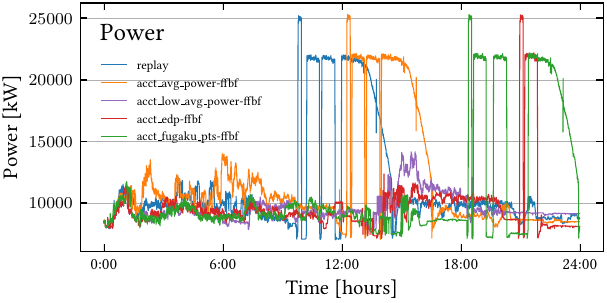}
    \caption{Studying the effects of incentive structures by using account information for prioritization. An Account's priority is based on the accumulated job behavior in the replay case (blue). Reprioritization based on this behavior is shown based on 
    descending average power (orange), ascending  average power (purple), \gls{edp} (red),
    and Fugaku points (green).
    }
    \label{fig:FgkPts-to-Frontier}
    \vspace{-0.2in}
\end{figure}

In Fig.~\ref{fig:FgkPts-to-Frontier}, we show the resulting power plots when applying different redeeming mechanisms to the same day as in Fig.~\ref{fig:Frontier0118}.
For the collection phase the replay policy (blue) was used, to illustrate the example.
The prioritization for the policies is based on an account's previous behavior on:
\textit{average power} (orange, higher is better),
\textit{low average power} (purple, lower is better),
\textit{\gls{edp}} (red),
\textit{Fugaku points} (green).
Fugaku points reward low average energy consumption in the collection phase.
The high power demand of the three large jobs in the collection was not rewarded in the redeeming phase (green), while the generally low power profile was rewarded as intended in \cite{Solorzano2024}.
This example illustrates how \gls{S-RAPS} can be used to run what-if studies that are difficult to realize on production systems.


\subsection{Use-Case: Using ML for scheduling decisions}
\label{sec:Evaluation:ML}
To demonstrate \gls{S-RAPS} utility, we evaluate a new \acrfull{ML} guided scheduling policy prototyped using \gls{S-RAPS}.

\begin{figure}[t]
    \centering    \centerline
{\includegraphics[width=0.9\columnwidth]{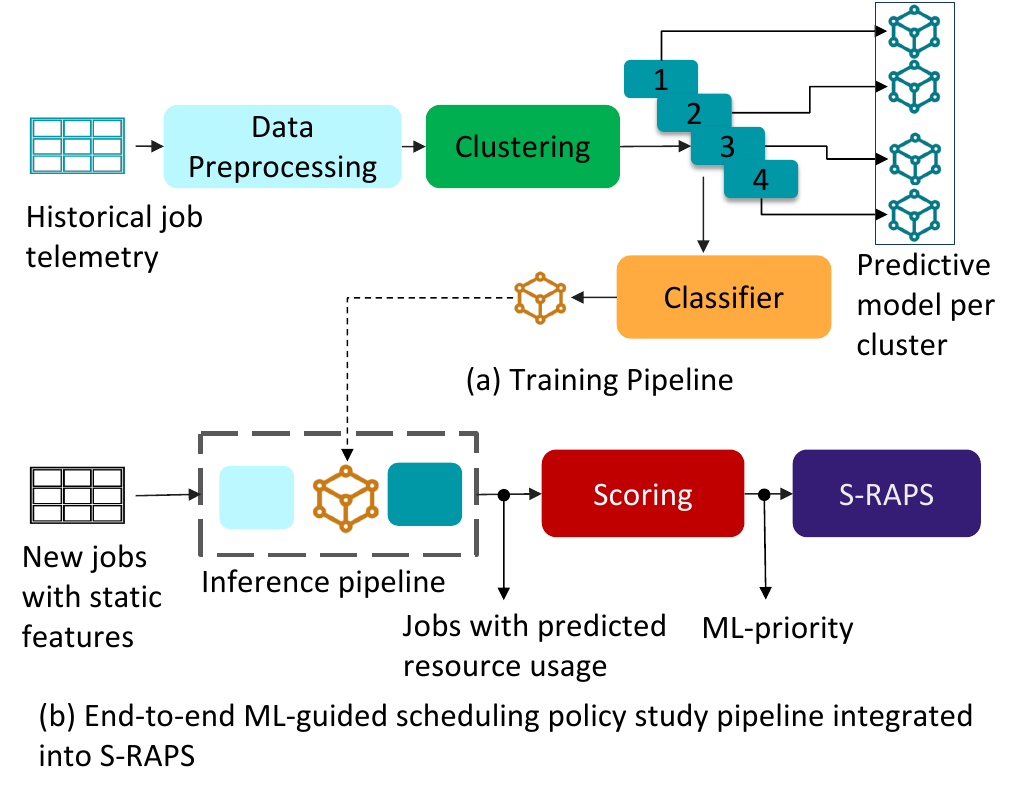}}
\vspace{-0.2in}
  \caption{Overview of 
  the ML-guided scheduling pipeline. 
  }
    \label{fig:methods}
    \vspace{-0.2in}
\end{figure}

\subsubsection{Training Phase:}
\textbf{(1) Clustering.} We partition historical jobs into behavioral clusters using both static (e.g., job size) and dynamic (e.g., power traces) features
using K-means
clustering.
\textbf{(2) Classification.} Since dynamic features are unavailable at submission, we train a Random Forest
model to learn the relationships between job characteristics (using pre-submission features) and the target metric.
This enables real-time mapping during inference time, without requiring telemetry for new jobs.
\textbf{(3) Prediction.} For each cluster, we train a model to predict target metrics such as runtime, power, memory -- based on static inputs. 

\subsubsection{Inference Phase:}
Upon job submission, we normalizes static features, predict the cluster label, invoke the corresponding model, and estimate performance. This design avoids global approximations and ensures predictions are tied to the job’s class.

Jobs are ranked via a score computed from predicted metrics and selected static features with the equation:

$S({X_i})~=~\sum_{j=1}^{K}{\alpha_j}~\cdot~\exp$\begin{small}${\left({\sqrt{{X^j_i}~+~1}}^{-1}\right)}$\end{small}.

\noindent Where $S(X_i)$ denotes the score of job $X_i$.
$X^{j}_{i}$ denotes the $j$-th feature of the $i$-th job.
$\alpha_j$ is the coefficient of feature $j$.
The exponential function captures fine-grained differences, allowing prioritization based on predicted system-level impact. 
Unlike single-objective schedulers, this supports trade-offs across throughput, wait time, turnaround, and energy.
\subsubsection{Evaluation:}
Some datasets (e.g., PM100) report time-series metrics, while others provide scalar summaries. 
Since timeseries data is inherently noisy and high-dimensional,
this causes inaccuracies in the clustering.
Hence, we extract summary statistics from timeseries metrics such as maximum, minimum, and standard deviation to retain key behavioral patterns.
\begin{figure}[t]
    \centering
    \subfigure[Power Consumption vs timestep for Fugaku]{
        \includegraphics[width=0.9\columnwidth]{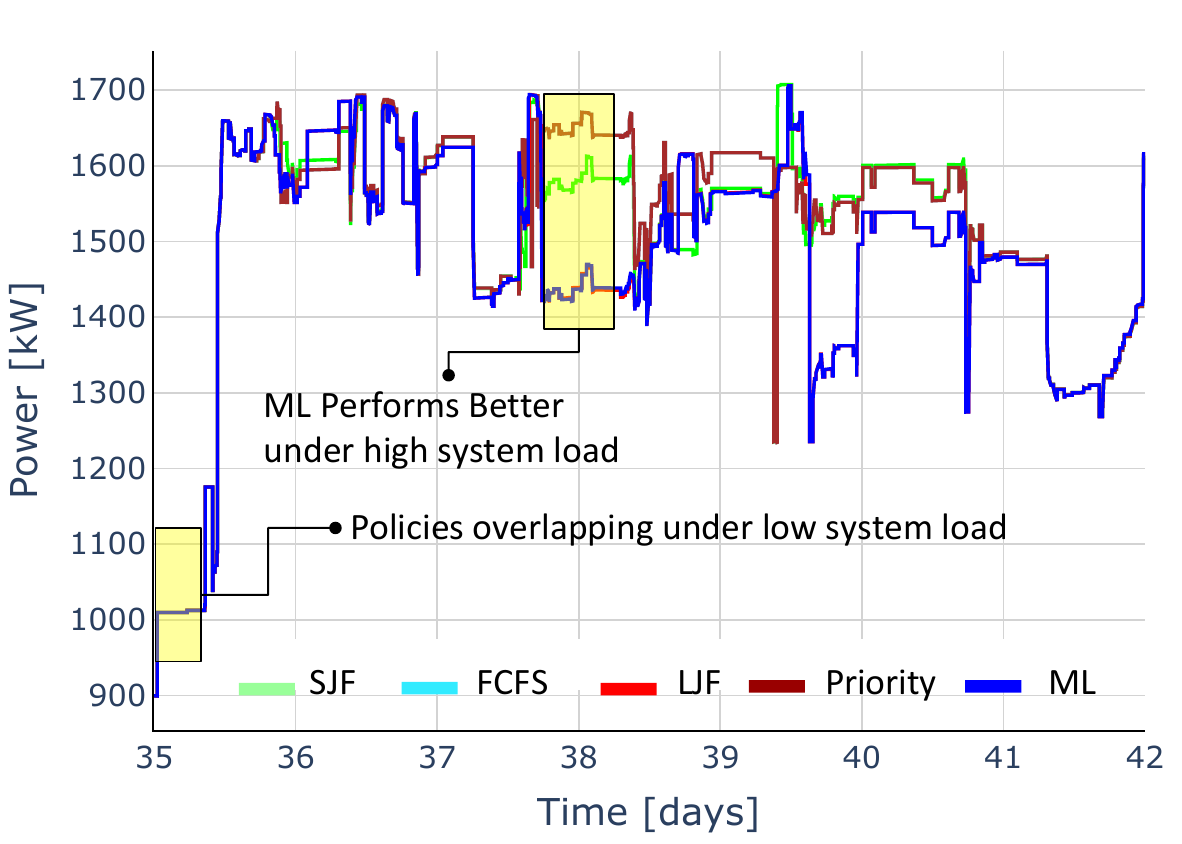}
        \label{fig:fugaku-line}
    }
    \hfill
    \subfigure[L2-Normalized multi-objective comparison among policies for Fugaku]{
        \includegraphics[width=0.75\columnwidth]{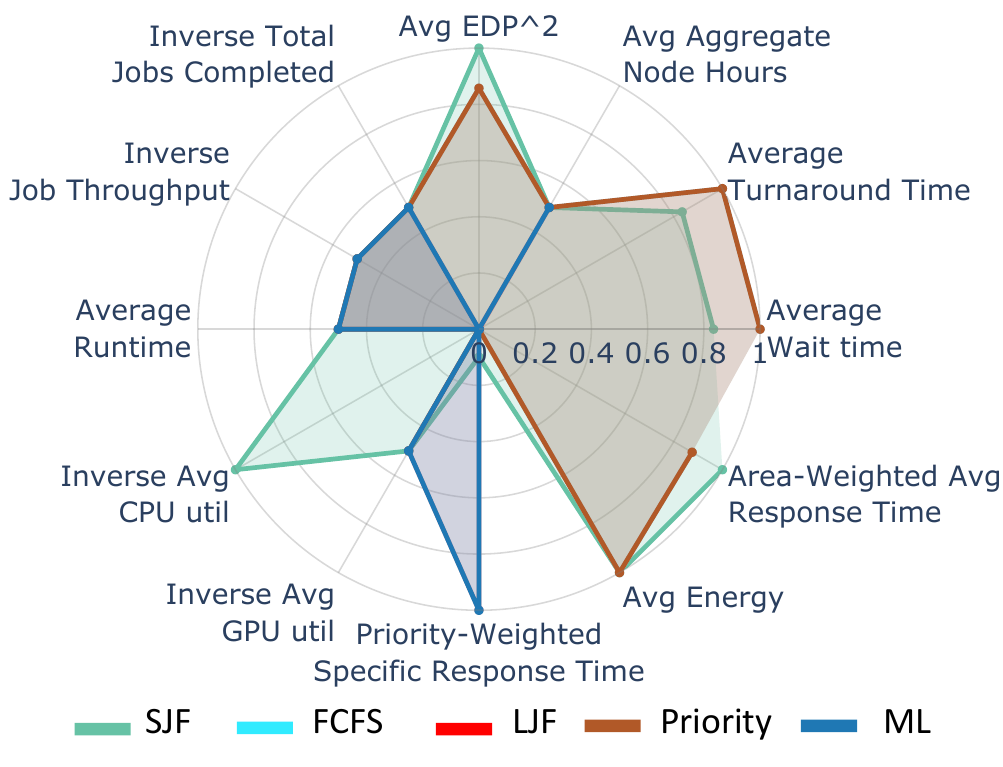}
        \label{fig:fugaku-radar}
    }
    \caption{Comparison of various scheduling policies. (a) Under high system load, ML-guided 
    policy yields 
    lower power per timestep. and (b) It also improves overall 
    system efficiency across multiple metrics (lower is better).}
\label{fig:radar-comparison}
    \vspace{-0.2in}
\end{figure}

In the Fugaku dataset, under low system load (16\% requested node utilization), as observed in the left yellow-marked region in Fig.~\ref{fig:fugaku-line}, all scheduling policies exhibit similar behavior. 
This happens because with abundance in resources, most jobs are scheduled immediately, resulting in minimal queuing delay and limited influence of scheduling policy. 
In contrast, under high load, right yellow-marked region in Fig.~\ref{fig:fugaku-line}, when aggregate job demand exceeds available nodes, 
the ML-guided policy reduces power spikes per timestep by prioritizing smaller jobs over larger ones.

To evaluate the broader impact of scheduling policies on the overall efficiency of the system, we analyze the policies for a time window with higher resource constraints and variable job sizes.
Figure~\ref{fig:fugaku-radar} shows consistent trends across datasets: (1) the ML-guided policy achieves the best trade-off across multiple objectives--including lower average wait time, turnaround, and energy consumption, 
compared to the baseline. (2) Under high load, ML-guided policy consistently yields the lowest job turnaround time and energy-delay product, increasing science per energy spent. 
\section{Discussion and Future Work}
\label{sec:Discussion}
The presented capabilities show a novel way of utilizing operational data --- now generally collect during operation --- to better understand and predict the behavior of our systems. 
This work allows studying scenarios and extensions for system- and simulation-needs 
expanding potential use-cases.
%
We enable integration of arbitrary scheduler and job-trace datasets and allow
replay and rescheduling to study their characteristics and exercise what-if scenarios.
We are able to seamlessly interact with system specific cooling models which can be easily generated~\cite{Greenwood2024}, while the power simulation is not a mere aggregation of synchronized trace information, but an accurate computation of component behavior~\cite{Wojda2024}.
%
The work presented allows any user to model their system with the existing tools, and study use-cases based user behavior, job-mix, and the scheduler, extending beyond system, cooling, and location.

The examples show this clearly: 
the use-cases presented in Sec.~\ref{sec:Evaluation:Dataloaders} and Sec~\ref{sec:Evaluation:extSchedulers} demonstrate what impact scheduling makes on the power response of a system. 
Sec.~\ref{sec:Evaluation:ML} shows how \gls{S-RAPS} allowed to prototype new algorithms, potentially avoiding averse effects.

For future work, identified gaps are the current need of job traces, and power profiles. 
Figure~\ref{fig:Adastra15days} showed that with perfect information of the job profile, we can accurately predict the systems power swings. However, if this information is not available, we have to rely on user estimates, or fingerprinting and prediction, which are prime candidates for future work.

\section{Conclusion}
\label{sec:Conclusion}

This work provides the first data center digital twin extended for scheduling.
We show how we build on a community-driven approach, with the \gls{exadigit} effort, and integrate scheduling capability into this interacting systems of simulators.
We present \acrfull{S-RAPS} the scheduler extension of \gls{exadigit}'s \gls{RAPS} and show how we can study scheduler-induced what-if scenarios and observe the connected digital twin simulations.

This is a first-of-its-kind study, which demonstrates significant improvements over the state of the art, by enabling other schedulers to tie into \glspl{DCDT} via \gls{S-RAPS}, as shown for the two external scheduling simulators, \gls{ScheduleFlow} and \gls{FastSim}.
Finally, we provided case-studies on how the work is used to simulate a scheduling power dip using FastSim, we showed how \gls{S-RAPS} can be used to study incentive structures for schedulers and showed how ML-guided schedulers can be studied using \gls{S-RAPS} as interface to \acrlongpl{DT}.

\footnotesize{
\begin{acks}
This research used resources of the Oak Ridge Leadership Computing Facility at the Oak Ridge National Laboratory, which is supported by the Office of Science of the U.S. Department of Energy under Contract No. DE-AC05-00OR22725.
Part of this work was performed under the auspices of the U.S.~Department of Energy by Lawrence Livermore National Laboratory under Contract DE-AC52-07NA27344 and was supported by the LLNL-LDRD Program under Project No. 24-SI-005 (LLNL-CONF-2004842).
This material is based upon work supported by the U.S. Department of Energy, Office of Science under Award Number DE-SC0022843 (ECRP).
Part of this work was authored in part by the National Renewable Energy Laboratory for the U.S. Department of Energy (DOE) under Contract No. DE-AC36-08GO28308.
\end{acks}
}
\bibliographystyle{ACM-Reference-Format}
\bibliography{main.bib}


\begin{thebibliography}{42}


\ifx \showCODEN    \undefined \def \showCODEN     #1{\unskip}     \fi
\ifx \showDOI      \undefined \def \showDOI       #1{#1}\fi
\ifx \showISBNx    \undefined \def \showISBNx     #1{\unskip}     \fi
\ifx \showISBNxiii \undefined \def \showISBNxiii  #1{\unskip}     \fi
\ifx \showISSN     \undefined \def \showISSN      #1{\unskip}     \fi
\ifx \showLCCN     \undefined \def \showLCCN      #1{\unskip}     \fi
\ifx \shownote     \undefined \def \shownote      #1{#1}          \fi
\ifx \showarticletitle \undefined \def \showarticletitle #1{#1}   \fi
\ifx \showURL      \undefined \def \showURL       {\relax}        \fi
\providecommand\bibfield[2]{#2}
\providecommand\bibinfo[2]{#2}
\providecommand\natexlab[1]{#1}
\providecommand\showeprint[2][]{arXiv:#2}

\bibitem[Adamson et~al\mbox{.}(2023)]%
        {Adamson2023}
\bibfield{author}{\bibinfo{person}{Ryan Adamson}, \bibinfo{person}{Tim
  Osborne}, \bibinfo{person}{Corwin Lester}, {and} \bibinfo{person}{Rachel
  Palumbo}.} \bibinfo{year}{2023}\natexlab{}.
\newblock \showarticletitle{STREAM: A Scalable Federated HPC Telemetry
  Platform}. In \bibinfo{booktitle}{\emph{Proceedings of the Cray User Group
  2023}}. Oak Ridge National Laboratory (ORNL), Oak Ridge, TN (United States).
\newblock
\urldef\tempurl%
\url{https://www.osti.gov/biblio/1995656}
\showURL{%
\tempurl}


\bibitem[Alaoui~Ismaili et~al\mbox{.}(2023)]%
        {alaoui2023porting}
\bibfield{author}{\bibinfo{person}{Naima Alaoui~Ismaili},
  \bibinfo{person}{Philippe Wautelet}, \bibinfo{person}{Juan Escobar~Munoz},
  {and} \bibinfo{person}{Gabriel Hautreux}.} \bibinfo{year}{2023}\natexlab{}.
\newblock \showarticletitle{Porting and optimizing Meso-NH to AMD MI250X GPUs}.
  In \bibinfo{booktitle}{\emph{Proceedings of the SC'23 Workshops of the
  International Conference on High Performance Computing, Network, Storage, and
  Analysis}}. \bibinfo{pages}{1900--1905}.
\newblock


\bibitem[Allcock et~al\mbox{.}(2017)]%
        {Allcock2017}
\bibfield{author}{\bibinfo{person}{William Allcock}, \bibinfo{person}{Paul
  Rich}, \bibinfo{person}{Yuping Fan}, {and} \bibinfo{person}{Zhiling Lan}.}
  \bibinfo{year}{2017}\natexlab{}.
\newblock \showarticletitle{Experience and Practice of Batch Scheduling on
  Leadership Supercomputers at Argonne}. In \bibinfo{booktitle}{\emph{21st
  Workshop on Job Scheduling Strategies for Parallel Processing held in
  conjunction with IPDPS 2017 (Orlando, FL, US, 06/02/2017 - 06/02/2017)}}.
  \bibinfo{publisher}{Springer}, \bibinfo{address}{Orlando, Florida},
  \bibinfo{pages}{1 -- 24,}.
\newblock
\urldef\tempurl%
\url{https://doi.org/10.1007/978-3-319-77398-8_1}
\showDOI{\tempurl}


\bibitem[Antici et~al\mbox{.}(2024)]%
        {antici2024fdata}
\bibfield{author}{\bibinfo{person}{Francesco Antici}, \bibinfo{person}{Andrea
  Bartolini}, \bibinfo{person}{Jens Domke}, \bibinfo{person}{Zeynep Kiziltan},
  {and} \bibinfo{person}{Keiji Yamamoto}.} \bibinfo{year}{2024}\natexlab{}.
\newblock \bibinfo{title}{{F-DATA}: {A} {Fugaku} {Workload} {Dataset} for
  {Job}-centric {Predictive} {Modelling} in {HPC} {Systems}}.
\newblock
\newblock
\urldef\tempurl%
\url{https://doi.org/10.5281/zenodo.11467483}
\showDOI{\tempurl}


\bibitem[Antici et~al\mbox{.}(2023)]%
        {antici2023pm100}
\bibfield{author}{\bibinfo{person}{Francesco Antici}, \bibinfo{person}{Mohsen
  Seyedkazemi~Ardebili}, \bibinfo{person}{Andrea Bartolini}, {and}
  \bibinfo{person}{Zeynep Kiziltan}.} \bibinfo{year}{2023}\natexlab{}.
\newblock \showarticletitle{PM100: A job power consumption dataset of a
  large-scale production HPC system}. In \bibinfo{booktitle}{\emph{Proceedings
  of the SC'23 Workshops of the International Conference on High Performance
  Computing, Network, Storage, and Analysis}}. \bibinfo{publisher}{Zenodo},
  \bibinfo{address}{Denver, CO}, \bibinfo{pages}{1812--1819}.
\newblock


\bibitem[Athavale et~al\mbox{.}(2024)]%
        {athavale2024digital}
\bibfield{author}{\bibinfo{person}{Jyotika Athavale}, \bibinfo{person}{Cullen
  Bash}, \bibinfo{person}{Wesley Brewer}, \bibinfo{person}{Matthias Maiterth},
  \bibinfo{person}{Dejan Milojicic}, \bibinfo{person}{Harry Petty}, {and}
  \bibinfo{person}{Soumyendu Sarkar}.} \bibinfo{year}{2024}\natexlab{}.
\newblock \showarticletitle{Digital Twins for Data Centers}.
\newblock \bibinfo{journal}{\emph{Computer}} \bibinfo{volume}{57},
  \bibinfo{number}{10} (\bibinfo{year}{2024}), \bibinfo{pages}{151--158}.
\newblock
\urldef\tempurl%
\url{https://doi.org/10.1109/MC.2024.3436945}
\showDOI{\tempurl}


\bibitem[Bo\"{e}zennec et~al\mbox{.}(2024)]%
        {Boezennec2024}
\bibfield{author}{\bibinfo{person}{Robin Bo\"{e}zennec}, \bibinfo{person}{Fanny
  Dufoss\'{e}}, {and} \bibinfo{person}{Guillaume Pallez}.}
  \bibinfo{year}{2024}\natexlab{}.
\newblock \showarticletitle{Qualitatively Analyzing Optimization Objectives in
  the Design of HPC Resource Manager}.
\newblock \bibinfo{journal}{\emph{ACM Trans. Model. Perform. Eval. Comput.
  Syst.}} \bibinfo{volume}{9}, \bibinfo{number}{4}, Article
  \bibinfo{articleno}{15} (\bibinfo{date}{Dec.} \bibinfo{year}{2024}),
  \bibinfo{numpages}{28}~pages.
\newblock
\showISSN{2376-3639}
\urldef\tempurl%
\url{https://doi.org/10.1145/3701986}
\showDOI{\tempurl}


\bibitem[Borghesi et~al\mbox{.}(2023)]%
        {borghesi2023m100}
\bibfield{author}{\bibinfo{person}{Andrea Borghesi}, \bibinfo{person}{Carmine
  Di~Santi}, \bibinfo{person}{Martin Molan},
  \bibinfo{person}{Mohsen~Seyedkazemi Ardebili}, \bibinfo{person}{Alessio
  Mauri}, \bibinfo{person}{Massimiliano Guarrasi}, \bibinfo{person}{Daniela
  Galetti}, \bibinfo{person}{Mirko Cestari}, \bibinfo{person}{Francesco
  Barchi}, \bibinfo{person}{Luca Benini}, {et~al\mbox{.}}}
  \bibinfo{year}{2023}\natexlab{}.
\newblock \showarticletitle{M100 exadata: a data collection campaign on the
  cineca’s marconi100 tier-0 supercomputer}.
\newblock \bibinfo{journal}{\emph{Scientific Data}} \bibinfo{volume}{10},
  \bibinfo{number}{1} (\bibinfo{year}{2023}), \bibinfo{pages}{288}.
\newblock


\bibitem[Brewer et~al\mbox{.}(2024)]%
        {brewer2024digital}
\bibfield{author}{\bibinfo{person}{Wesley Brewer}, \bibinfo{person}{Matthias
  Maiterth}, \bibinfo{person}{Vineet Kumar}, \bibinfo{person}{Rafal Wojda},
  \bibinfo{person}{Sedrick Bouknight}, \bibinfo{person}{Jesse Hines},
  \bibinfo{person}{Woong Shin}, \bibinfo{person}{Scott Greenwood},
  \bibinfo{person}{David Grant}, \bibinfo{person}{Wesley Williams}, {and}
  \bibinfo{person}{Feiyi Wang}.} \bibinfo{year}{2024}\natexlab{}.
\newblock \showarticletitle{A Digital Twin Framework for Liquid-cooled
  Supercomputers as Demonstrated at Exascale}. In
  \bibinfo{booktitle}{\emph{Proceedings of the International Conference for
  High Performance Computing, Networking, Storage, and Analysis}}
  \emph{(\bibinfo{series}{SC '24})}. \bibinfo{publisher}{IEEE Press},
  \bibinfo{address}{Atlanta, GA, USA}, Article \bibinfo{articleno}{23},
  \bibinfo{numpages}{18}~pages.
\newblock
\showISBNx{9798350352917}
\urldef\tempurl%
\url{https://doi.org/10.1109/SC41406.2024.00029}
\showDOI{\tempurl}


\bibitem[Brewer et~al\mbox{.}(2025)]%
        {S-RAPSgit}
\bibfield{author}{\bibinfo{person}{Wesley Brewer}, \bibinfo{person}{Rafal
  Wojda}, \bibinfo{person}{Matthias Maiterth}, \bibinfo{person}{Sedrick
  Bouknight}, \bibinfo{person}{Jesse Hines}, \bibinfo{person}{Jake Webb},
  \bibinfo{person}{Rashadul Kabir}, \bibinfo{person}{Bertrand Cirou}, {and}
  \bibinfo{person}{Kevin Menear}.} \bibinfo{year}{2025}\natexlab{}.
\newblock \bibinfo{title}{exadigit/RAPS -- S-RAPS branch}.
\newblock
  \bibinfo{howpublished}{\url{https://code.ornl.gov/exadigit/raps/-/tree/S-RAPS?ref_type=tags}}.
\newblock


\bibitem[Buyya and Murshed(2002)]%
        {Buyya2002gridsim}
\bibfield{author}{\bibinfo{person}{Rajkumar Buyya} {and}
  \bibinfo{person}{Manzur Murshed}.} \bibinfo{year}{2002}\natexlab{}.
\newblock \showarticletitle{Gridsim: A toolkit for the modeling and simulation
  of distributed resource management and scheduling for grid computing}.
\newblock \bibinfo{journal}{\emph{Concurrency and computation: practice and
  experience}} \bibinfo{volume}{14}, \bibinfo{number}{13-15}
  (\bibinfo{year}{2002}), \bibinfo{pages}{1175--1220}.
\newblock


\bibitem[Casanova et~al\mbox{.}(2014)]%
        {Casanova2014SimGrid}
\bibfield{author}{\bibinfo{person}{Henri Casanova}, \bibinfo{person}{Arnaud
  Giersch}, \bibinfo{person}{Arnaud Legrand}, \bibinfo{person}{Martin Quinson},
  {and} \bibinfo{person}{Frédéric Suter}.} \bibinfo{year}{2014}\natexlab{}.
\newblock \showarticletitle{Versatile, scalable, and accurate simulation of
  distributed applications and platforms}.
\newblock \bibinfo{journal}{\emph{J. Parallel and Distrib. Comput.}}
  \bibinfo{volume}{74}, \bibinfo{number}{10} (\bibinfo{year}{2014}),
  \bibinfo{pages}{2899--2917}.
\newblock
\showISSN{0743-7315}
\urldef\tempurl%
\url{https://doi.org/10.1016/j.jpdc.2014.06.008}
\showDOI{\tempurl}


\bibitem[Chapin et~al\mbox{.}(1999)]%
        {Chapin1999}
\bibfield{author}{\bibinfo{person}{Steve~J. Chapin}, \bibinfo{person}{Walfredo
  Cirne}, \bibinfo{person}{Dror~G. Feitelson}, \bibinfo{person}{James~Patton
  Jones}, \bibinfo{person}{Scott~T. Leutenegger}, \bibinfo{person}{Uwe
  Schwiegelshohn}, \bibinfo{person}{Warren Smith}, {and} \bibinfo{person}{David
  Talby}.} \bibinfo{year}{1999}\natexlab{}.
\newblock \showarticletitle{Benchmarks and Standards for the Evaluation of
  Parallel Job Schedulers}. In \bibinfo{booktitle}{\emph{Job Scheduling
  Strategies for Parallel Processing}},
  \bibfield{editor}{\bibinfo{person}{Dror~G. Feitelson} {and}
  \bibinfo{person}{Larry Rudolph}} (Eds.). \bibinfo{publisher}{Springer Berlin
  Heidelberg}, \bibinfo{address}{Berlin, Heidelberg}, \bibinfo{pages}{67--90}.
\newblock
\showISBNx{978-3-540-47954-3}


\bibitem[Cirou(2024)]%
        {Adastra15D}
\bibfield{author}{\bibinfo{person}{Cirou}.} \bibinfo{year}{2024}\natexlab{}.
\newblock \bibinfo{booktitle}{\emph{Adastra jobs MI250 15days}}.
\newblock
\urldef\tempurl%
\url{https://doi.org/10.5281/zenodo.14007065}
\showDOI{\tempurl}


\bibitem[Dutot et~al\mbox{.}(2016)]%
        {Dutot2016}
\bibfield{author}{\bibinfo{person}{Pierre-Fran{\c c}ois Dutot},
  \bibinfo{person}{Michael Mercier}, \bibinfo{person}{Millian Poquet}, {and}
  \bibinfo{person}{Olivier Richard}.} \bibinfo{year}{2016}\natexlab{}.
\newblock \showarticletitle{{Batsim: a Realistic Language-Independent Resources
  and Jobs Management Systems Simulator}}. In \bibinfo{booktitle}{\emph{{20th
  Workshop on Job Scheduling Strategies for Parallel Processing (JSSPP)}}}.
  \bibinfo{address}{Chicago, United States}.
\newblock
\urldef\tempurl%
\url{https://hal.science/hal-01333471}
\showURL{%
\tempurl}


\bibitem[Facility({[n.\,d.]})]%
        {FrontierUserGuide}
\bibfield{author}{\bibinfo{person}{Oak Ridge Leadership~Computing Facility}.}
  \bibinfo{year}{[n.\,d.]}\natexlab{}.
\newblock \bibinfo{title}{Frontier User Guide}.
\newblock
\newblock
\urldef\tempurl%
\url{https://docs.olcf.ornl.gov/systems/frontier_user_guide.html}
\showURL{%
\tempurl}


\bibitem[Fan et~al\mbox{.}(2019)]%
        {Fan2019BBSched}
\bibfield{author}{\bibinfo{person}{Yuping Fan}, \bibinfo{person}{Zhiling Lan},
  \bibinfo{person}{Paul Rich}, \bibinfo{person}{William~E. Allcock},
  \bibinfo{person}{Michael~E. Papka}, \bibinfo{person}{Brian Austin}, {and}
  \bibinfo{person}{David Paul}.} \bibinfo{year}{2019}\natexlab{}.
\newblock \showarticletitle{Scheduling Beyond CPUs for HPC}. In
  \bibinfo{booktitle}{\emph{Proceedings of the 28th International Symposium on
  High-Performance Parallel and Distributed Computing}} (Phoenix, AZ, USA)
  \emph{(\bibinfo{series}{HPDC '19})}. \bibinfo{publisher}{Association for
  Computing Machinery}, \bibinfo{address}{New York, NY, USA},
  \bibinfo{pages}{97–108}.
\newblock
\showISBNx{9781450366700}
\urldef\tempurl%
\url{https://doi.org/10.1145/3307681.3325401}
\showDOI{\tempurl}


\bibitem[Gainaru et~al\mbox{.}(2019)]%
        {Gainaru2019}
\bibfield{author}{\bibinfo{person}{Ana Gainaru}, \bibinfo{person}{Hongyang
  Sun}, \bibinfo{person}{Guillaume Aupy}, \bibinfo{person}{Yuankai Huo},
  \bibinfo{person}{Bennett~A Landman}, {and} \bibinfo{person}{Padma Raghavan}.}
  \bibinfo{year}{2019}\natexlab{}.
\newblock \showarticletitle{On-the-fly scheduling versus reservation-based
  scheduling for unpredictable workflows}.
\newblock \bibinfo{journal}{\emph{The International Journal of High Performance
  Computing Applications}} \bibinfo{volume}{33}, \bibinfo{number}{6}
  (\bibinfo{year}{2019}), \bibinfo{pages}{1140--1158}.
\newblock
\urldef\tempurl%
\url{https://doi.org/10.1177/1094342019841681}
\showDOI{\tempurl}
\showeprint{https://doi.org/10.1177/1094342019841681}


\bibitem[Galleguillos et~al\mbox{.}(2018)]%
        {Galleguillos2018}
\bibfield{author}{\bibinfo{person}{Cristian Galleguillos},
  \bibinfo{person}{Alina S{\^i}rbu}, \bibinfo{person}{Zeynep Kiziltan},
  \bibinfo{person}{Ozalp Babaoglu}, \bibinfo{person}{Andrea Borghesi}, {and}
  \bibinfo{person}{Thomas Bridi}.} \bibinfo{year}{2018}\natexlab{}.
\newblock \showarticletitle{Data-Driven Job Dispatching in HPC Systems}. In
  \bibinfo{booktitle}{\emph{Machine Learning, Optimization, and Big Data}},
  \bibfield{editor}{\bibinfo{person}{Giuseppe Nicosia}, \bibinfo{person}{Panos
  Pardalos}, \bibinfo{person}{Giovanni Giuffrida}, {and}
  \bibinfo{person}{Renato Umeton}} (Eds.). \bibinfo{publisher}{Springer
  International Publishing}, \bibinfo{address}{Cham},
  \bibinfo{pages}{449--461}.
\newblock
\showISBNx{978-3-319-72926-8}


\bibitem[Gay and Caniou(2006)]%
        {gay2006simbatch}
\bibfield{author}{\bibinfo{person}{Jean-S{\'e}bastien Gay} {and}
  \bibinfo{person}{Yves Caniou}.} \bibinfo{year}{2006}\natexlab{}.
\newblock \bibinfo{title}{Simbatch: an API for simulating and predicting the
  performance of parallel resources and batch systems}.
\newblock
\newblock
\newblock
\shownote{Research Report}.


\bibitem[Goponenko et~al\mbox{.}(2022)]%
        {goponenko2022metrics}
\bibfield{author}{\bibinfo{person}{Alexander~V. Goponenko},
  \bibinfo{person}{Kenneth Lamar}, \bibinfo{person}{Christina Peterson},
  \bibinfo{person}{Benjamin~A. Allan}, \bibinfo{person}{Jim~M. Brandt}, {and}
  \bibinfo{person}{Damian Dechev}.} \bibinfo{year}{2022}\natexlab{}.
\newblock \showarticletitle{Metrics for Packing Efficiency and Fairness of HPC
  Cluster Batch Job Scheduling}. In \bibinfo{booktitle}{\emph{2022 IEEE 34th
  International Symposium on Computer Architecture and High Performance
  Computing (SBAC-PAD)}}. \bibinfo{publisher}{IEEE},
  \bibinfo{address}{Bordeaux, France}, \bibinfo{pages}{241--252}.
\newblock
\urldef\tempurl%
\url{https://doi.org/10.1109/SBAC-PAD55451.2022.00035}
\showDOI{\tempurl}


\bibitem[Greenwood et~al\mbox{.}(2024)]%
        {Greenwood2024}
\bibfield{author}{\bibinfo{person}{S. Greenwood}, \bibinfo{person}{V. Kumar},
  {and} \bibinfo{person}{W. Brewer}.} \bibinfo{year}{2024}\natexlab{}.
\newblock \showarticletitle{Thermo-fluid Modeling Framework for Supercomputer
  Digital Twins: Part 2, Automated Cooling Models}. In
  \bibinfo{booktitle}{\emph{America Modelica Conference}}. Modelica
  Association, \bibinfo{pages}{210--219}.
\newblock


\bibitem[Jackson et~al\mbox{.}(2001)]%
        {Jackson2001Maui}
\bibfield{author}{\bibinfo{person}{David Jackson}, \bibinfo{person}{Quinn
  Snell}, {and} \bibinfo{person}{Mark Clement}.}
  \bibinfo{year}{2001}\natexlab{}.
\newblock \showarticletitle{Core Algorithms of the Maui Scheduler}. In
  \bibinfo{booktitle}{\emph{Job Scheduling Strategies for Parallel
  Processing}}. \bibinfo{publisher}{Springer Berlin Heidelberg},
  \bibinfo{address}{Berlin, Heidelberg}, \bibinfo{pages}{87--102}.
\newblock
\showISBNx{978-3-540-45540-0}


\bibitem[Klus{\'a}{\v{c}}ek et~al\mbox{.}(2019)]%
        {Klusavcek2019alea}
\bibfield{author}{\bibinfo{person}{Dalibor Klus{\'a}{\v{c}}ek},
  \bibinfo{person}{Mehmet Soysal}, {and} \bibinfo{person}{Fr{\'e}d{\'e}ric
  Suter}.} \bibinfo{year}{2019}\natexlab{}.
\newblock \showarticletitle{Alea--complex job scheduling simulator}. In
  \bibinfo{booktitle}{\emph{International Conference on Parallel Processing and
  Applied Mathematics}}. \bibinfo{publisher}{Springer},
  \bibinfo{address}{Bialystok, Poland}, \bibinfo{pages}{217--229}.
\newblock


\bibitem[Kumar et~al\mbox{.}(2024)]%
        {Kumar2024}
\bibfield{author}{\bibinfo{person}{Vineet Kumar}, \bibinfo{person}{Scott
  Greenwood}, \bibinfo{person}{Wes Brewer}, \bibinfo{person}{Wesley Williams},
  \bibinfo{person}{David Grant}, {and} \bibinfo{person}{Nathan Parkison}.}
  \bibinfo{year}{2024}\natexlab{}.
\newblock \showarticletitle{Thermo-Fluid Modeling Framework for Supercomputer
  Digital Twins: Part 1, Demonstration at Exascale}. In
  \bibinfo{booktitle}{\emph{PROCEEDINGS OF THE AMERICAN MODELICA CONFERENCE
  2024}}. Oak Ridge National Laboratory (ORNL), Oak Ridge, TN (United States).
\newblock
\urldef\tempurl%
\url{https://www.osti.gov/biblio/2480044}
\showURL{%
\tempurl}


\bibitem[{Lawrence Livermore National Laboratory}(2024)]%
        {last2024}
\bibfield{author}{\bibinfo{person}{{Lawrence Livermore National Laboratory}}.}
  \bibinfo{year}{2024}\natexlab{}.
\newblock \bibinfo{title}{Livermore Archive for System Telemetry (LAST)}.
\newblock
\newblock
\urldef\tempurl%
\url{https://github.com/LLNL/LAST}
\showURL{%
\tempurl}


\bibitem[Patki et~al\mbox{.}(2021)]%
        {patki2021monitoring}
\bibfield{author}{\bibinfo{person}{Tapasya Patki}, \bibinfo{person}{Adam
  Bertsch}, \bibinfo{person}{Ian Karlin}, \bibinfo{person}{Dong~H Ahn},
  \bibinfo{person}{Brian Van~Essen}, \bibinfo{person}{Barry Rountree},
  \bibinfo{person}{Bronis~R de Supinski}, {and} \bibinfo{person}{Nathan
  Besaw}.} \bibinfo{year}{2021}\natexlab{}.
\newblock \showarticletitle{Monitoring large scale supercomputers: A case study
  with the lassen supercomputer}. In \bibinfo{booktitle}{\emph{2021 IEEE
  International Conference on Cluster Computing (CLUSTER)}}.
  \bibinfo{publisher}{IEEE}, \bibinfo{address}{Portland, OR, United States},
  \bibinfo{pages}{468--480}.
\newblock


\bibitem[Patki et~al\mbox{.}(2019)]%
        {patki2019comparing}
\bibfield{author}{\bibinfo{person}{Tapasya Patki}, \bibinfo{person}{Zachary
  Frye}, \bibinfo{person}{Harsh Bhatia}, \bibinfo{person}{Francesco Di~Natale},
  \bibinfo{person}{James Glosli}, \bibinfo{person}{Helgi Ingolfsson}, {and}
  \bibinfo{person}{Barry Rountree}.} \bibinfo{year}{2019}\natexlab{}.
\newblock \showarticletitle{Comparing gpu power and frequency capping: A case
  study with the mummi workflow}. In \bibinfo{booktitle}{\emph{2019 IEEE/ACM
  Workflows in Support of Large-Scale Science (WORKS)}}. IEEE,
  \bibinfo{pages}{31--39}.
\newblock


\bibitem[Potts and Kovalyov(2000)]%
        {Potts2000}
\bibfield{author}{\bibinfo{person}{Chris~N. Potts} {and}
  \bibinfo{person}{Mikhail~Y. Kovalyov}.} \bibinfo{year}{2000}\natexlab{}.
\newblock \showarticletitle{Scheduling with batching: A review}.
\newblock \bibinfo{journal}{\emph{European Journal of Operational Research}}
  \bibinfo{volume}{120}, \bibinfo{number}{2} (\bibinfo{year}{2000}),
  \bibinfo{pages}{228--249}.
\newblock
\showISSN{0377-2217}
\urldef\tempurl%
\url{https://doi.org/10.1016/S0377-2217(99)00153-8}
\showDOI{\tempurl}


\bibitem[Ren et~al\mbox{.}(2017)]%
        {CQSim}
\bibfield{author}{\bibinfo{person}{Dongxu Ren}, \bibinfo{person}{Wei Tang},
  \bibinfo{person}{Xu Yang}, \bibinfo{person}{Yuping Fan}, {and}
  \bibinfo{person}{Zhiling Lan}.} \bibinfo{year}{2017}\natexlab{}.
\newblock
\newblock
\urldef\tempurl%
\url{https://github.com/SPEAR-IIT/CQSim}
\showURL{%
\tempurl}


\bibitem[Rodrigo et~al\mbox{.}(2017)]%
        {Rodrigo2017scsf}
\bibfield{author}{\bibinfo{person}{Gonzalo~P Rodrigo}, \bibinfo{person}{Erik
  Elmroth}, \bibinfo{person}{Per-Olov {\"O}stberg}, {and}
  \bibinfo{person}{Lavanya Ramakrishnan}.} \bibinfo{year}{2017}\natexlab{}.
\newblock \showarticletitle{{ScSF}: A scheduling simulation framework}. In
  \bibinfo{booktitle}{\emph{Workshop on Job Scheduling Strategies for Parallel
  Processing}}. \bibinfo{publisher}{Springer}, \bibinfo{address}{Cham, DE},
  \bibinfo{pages}{152--173}.
\newblock


\bibitem[Schroeder et~al\mbox{.}(2016)]%
        {Schroeder2016}
\bibfield{author}{\bibinfo{person}{Greyce~N. Schroeder},
  \bibinfo{person}{Charles Steinmetz}, \bibinfo{person}{Carlos~E. Pereira},
  {and} \bibinfo{person}{Danubia~B. Espindola}.}
  \bibinfo{year}{2016}\natexlab{}.
\newblock \showarticletitle{Digital Twin Data Modeling with AutomationML and a
  Communication Methodology for Data Exchange}.
\newblock \bibinfo{journal}{\emph{IFAC-PapersOnLine}} \bibinfo{volume}{49},
  \bibinfo{number}{30} (\bibinfo{year}{2016}), \bibinfo{pages}{12--17}.
\newblock
\showISSN{2405-8963}
\urldef\tempurl%
\url{https://doi.org/10.1016/j.ifacol.2016.11.115}
\showDOI{\tempurl}
\newblock
\shownote{4th IFAC Symposium on Telematics Applications TA 2016}.


\bibitem[Simakov et~al\mbox{.}(2018a)]%
        {Simakov2018-2}
\bibfield{author}{\bibinfo{person}{Nikolay~A. Simakov},
  \bibinfo{person}{Robert~L. DeLeon}, \bibinfo{person}{Martins~D. Innus},
  \bibinfo{person}{Matthew~D. Jones}, \bibinfo{person}{Joseph~P. White},
  \bibinfo{person}{Steven~M. Gallo}, \bibinfo{person}{Abani~K. Patra}, {and}
  \bibinfo{person}{Thomas~R. Furlani}.} \bibinfo{year}{2018}\natexlab{a}.
\newblock \showarticletitle{Slurm Simulator: Improving Slurm Scheduler
  Performance on Large HPC systems by Utilization of Multiple Controllers and
  Node Sharing}. In \bibinfo{booktitle}{\emph{Proceedings of the Practice and
  Experience on Advanced Research Computing: Seamless Creativity}} (Pittsburgh,
  PA, USA) \emph{(\bibinfo{series}{PEARC '18})}.
  \bibinfo{publisher}{Association for Computing Machinery},
  \bibinfo{address}{New York, NY, USA}, Article \bibinfo{articleno}{25},
  \bibinfo{numpages}{8}~pages.
\newblock
\showISBNx{9781450364461}
\urldef\tempurl%
\url{https://doi.org/10.1145/3219104.3219111}
\showDOI{\tempurl}


\bibitem[Simakov et~al\mbox{.}(2022)]%
        {Simakov2022}
\bibfield{author}{\bibinfo{person}{Nikolay~A. Simakov},
  \bibinfo{person}{Robert~L. Deleon}, \bibinfo{person}{Yuqing Lin},
  \bibinfo{person}{Phillip~S. Hoffmann}, {and} \bibinfo{person}{William~R.
  Mathias}.} \bibinfo{year}{2022}\natexlab{}.
\newblock \showarticletitle{Developing Accurate Slurm Simulator}. In
  \bibinfo{booktitle}{\emph{Practice and Experience in Advanced Research
  Computing 2022: Revolutionary: Computing, Connections, You}} (Boston, MA,
  USA) \emph{(\bibinfo{series}{PEARC '22})}. \bibinfo{publisher}{Association
  for Computing Machinery}, \bibinfo{address}{New York, NY, USA}, Article
  \bibinfo{articleno}{59}, \bibinfo{numpages}{4}~pages.
\newblock
\showISBNx{9781450391610}
\urldef\tempurl%
\url{https://doi.org/10.1145/3491418.3535178}
\showDOI{\tempurl}


\bibitem[Simakov et~al\mbox{.}(2018b)]%
        {Simakov2018SlurmSim}
\bibfield{author}{\bibinfo{person}{Nikolay~A. Simakov},
  \bibinfo{person}{Martins~D. Innus}, \bibinfo{person}{Matthew~D. Jones},
  \bibinfo{person}{Robert~L. DeLeon}, \bibinfo{person}{Joseph~P. White},
  \bibinfo{person}{Steven~M. Gallo}, \bibinfo{person}{Abani~K. Patra}, {and}
  \bibinfo{person}{Thomas~R. Furlani}.} \bibinfo{year}{2018}\natexlab{b}.
\newblock \showarticletitle{A Slurm Simulator: Implementation and Parametric
  Analysis}. In \bibinfo{booktitle}{\emph{High Performance Computing Systems.
  Performance Modeling, Benchmarking, and Simulation}},
  \bibfield{editor}{\bibinfo{person}{Stephen Jarvis}, \bibinfo{person}{Steven
  Wright}, {and} \bibinfo{person}{Simon Hammond}} (Eds.).
  \bibinfo{publisher}{Springer International Publishing},
  \bibinfo{address}{Cham, Germany}, \bibinfo{pages}{197--217}.
\newblock
\showISBNx{978-3-319-72971-8}


\bibitem[Skovira et~al\mbox{.}(1996)]%
        {EASY}
\bibfield{author}{\bibinfo{person}{Joseph Skovira}, \bibinfo{person}{Waiman
  Chan}, \bibinfo{person}{Honbo Zhou}, {and} \bibinfo{person}{David~A. Lifka}.}
  \bibinfo{year}{1996}\natexlab{}.
\newblock \showarticletitle{The {EASY} - LoadLeveler {API} Project}. In
  \bibinfo{booktitle}{\emph{Job Scheduling Strategies for Parallel Processing,
  IPPS'96 Workshop, Honolulu, Haiwai, USA, April 16, 1996, Proceedings}}
  \emph{(\bibinfo{series}{Lecture Notes in Computer Science},
  Vol.~\bibinfo{volume}{1162})}, \bibfield{editor}{\bibinfo{person}{Dror~G.
  Feitelson} {and} \bibinfo{person}{Larry Rudolph}} (Eds.).
  \bibinfo{publisher}{Springer}, \bibinfo{pages}{41--47}.
\newblock
\urldef\tempurl%
\url{https://doi.org/10.1007/BFB0022286}
\showDOI{\tempurl}


\bibitem[Sol\'{o}rzano et~al\mbox{.}(2024)]%
        {Solorzano2024}
\bibfield{author}{\bibinfo{person}{Ana Luisa~Veroneze Sol\'{o}rzano},
  \bibinfo{person}{Kento Sato}, \bibinfo{person}{Keiji Yamamoto},
  \bibinfo{person}{Fumiyoshi Shoji}, \bibinfo{person}{Jim~M. Brandt},
  \bibinfo{person}{Benjamin Schwaller}, \bibinfo{person}{Sara~Petra Walton},
  \bibinfo{person}{Jennifer Green}, {and} \bibinfo{person}{Devesh Tiwari}.}
  \bibinfo{year}{2024}\natexlab{}.
\newblock \bibinfo{title}{Toward Sustainable HPC: In-Production Deployment of
  Incentive-Based Power Efficiency Mechanism on the Fugaku Supercomputer}.
\newblock , \bibinfo{numpages}{16}~pages.
\newblock
\showISBNx{9798350352917}
\urldef\tempurl%
\url{https://doi.org/10.1109/SC41406.2024.00030}
\showDOI{\tempurl}


\bibitem[Takefusa et~al\mbox{.}(1999)]%
        {Takefusa1999Bricks}
\bibfield{author}{\bibinfo{person}{Atsuko Takefusa}, \bibinfo{person}{Satoshi
  Matsuoka}, \bibinfo{person}{Kento Aida}, \bibinfo{person}{Hidemoto Nakada},
  {and} \bibinfo{person}{Umpei Nagashima}.} \bibinfo{year}{1999}\natexlab{}.
\newblock \showarticletitle{Overview of a Performance Evaluation System for
  Global Computing Scheduling Algorithms}. In
  \bibinfo{booktitle}{\emph{Proceedings of the 8th IEEE International Symposium
  on High Performance Distributed Computing}} \emph{(\bibinfo{series}{HPDC
  '99})}. \bibinfo{publisher}{IEEE Computer Society}, \bibinfo{address}{USA},
  \bibinfo{pages}{11}.
\newblock
\showISBNx{0769502873}


\bibitem[Tao et~al\mbox{.}(2019)]%
        {Tao2019DigitalTI}
\bibfield{author}{\bibinfo{person}{Fei Tao}, \bibinfo{person}{He Zhang},
  \bibinfo{person}{Ang Liu}, {and} \bibinfo{person}{Andrew Y.~C. Nee}.}
  \bibinfo{year}{2019}\natexlab{}.
\newblock \showarticletitle{Digital Twin in Industry: State-of-the-Art}.
\newblock \bibinfo{journal}{\emph{IEEE Transactions on Industrial Informatics}}
   \bibinfo{volume}{15} (\bibinfo{year}{2019}), \bibinfo{pages}{2405--2415}.
\newblock
\urldef\tempurl%
\url{https://api.semanticscholar.org/CorpusID:68170459}
\showURL{%
\tempurl}


\bibitem[Vispi~Karkaria and Chen(2025)]%
        {Karkaria2025}
\bibfield{author}{\bibinfo{person}{Yi-Ping~Chen Vispi~Karkaria, Ying-Kuan~Tsai}
  {and} \bibinfo{person}{Wei Chen}.} \bibinfo{year}{2025}\natexlab{}.
\newblock \showarticletitle{An optimization-centric review on integrating
  artificial intelligence and digital twin technologies in manufacturing}.
\newblock \bibinfo{journal}{\emph{Engineering Optimization}}
  \bibinfo{volume}{57}, \bibinfo{number}{1} (\bibinfo{year}{2025}),
  \bibinfo{pages}{161--207}.
\newblock
\urldef\tempurl%
\url{https://doi.org/10.1080/0305215X.2024.2434201}
\showDOI{\tempurl}


\bibitem[Wilkinson et~al\mbox{.}(2023)]%
        {Wilkinson2023FastSim}
\bibfield{author}{\bibinfo{person}{Alex Wilkinson}, \bibinfo{person}{Jess
  Jones}, \bibinfo{person}{Harvey Richardson}, \bibinfo{person}{Tim Dykes},
  {and} \bibinfo{person}{Utz-Uwe Haus}.} \bibinfo{year}{2023}\natexlab{}.
\newblock \showarticletitle{A Fast Simulator to Enable HPC Scheduling Strategy
  Comparisons}. In \bibinfo{booktitle}{\emph{High Performance Computing: ISC
  High Performance 2023 International Workshops, Hamburg, Germany, May 21–25,
  2023, Revised Selected Papers}}. \bibinfo{publisher}{Springer-Verlag},
  \bibinfo{address}{Berlin, Heidelberg}, \bibinfo{pages}{320–333}.
\newblock
\showISBNx{978-3-031-40842-7}
\urldef\tempurl%
\url{https://doi.org/10.1007/978-3-031-40843-4_24}
\showDOI{\tempurl}


\bibitem[Wojda et~al\mbox{.}(2024)]%
        {Wojda2024}
\bibfield{author}{\bibinfo{person}{Rafal~P. Wojda}, \bibinfo{person}{Matthias
  Maiterth}, \bibinfo{person}{Sedrick Bouknight}, {and} \bibinfo{person}{Wesley
  Brewer}.} \bibinfo{year}{2024}\natexlab{}.
\newblock \showarticletitle{Dynamic Modeling of Power Conversion Stages for an
  Exascale Supercomputer}. In \bibinfo{booktitle}{\emph{2024 IEEE Energy
  Conversion Congress and Exposition (ECCE)}}. \bibinfo{pages}{1595--1601}.
\newblock
\urldef\tempurl%
\url{https://doi.org/10.1109/ECCE55643.2024.10861715}
\showDOI{\tempurl}


\end{thebibliography}

\clearpage
\appendix
\twocolumn[%
{\begin{center}
\Huge
Appendix: Artifact Description/Artifact Evaluation
\end{center}}
]


\appendixAD

\section{Overview of Contributions and Artifacts}

\subsection{Paper's Main Contributions}


\begin{description}

\item[$C_1$] \textbf{\acrshort{S-RAPS}}: \gls{S-RAPS} is an extension to\\
\gls{exadigit}'s \gls{RAPS} providing functionality to integrate scheduling capabilities for the \gls{exadigit} \gls{DT} framework.

\item[$C_2$] \textbf{Dataloader extensions to \gls{S-RAPS}}: The dataloaders make it possible to load a open datasets into the \gls{exadigit} framework via \gls{S-RAPS} for simulation.

\item[$C_3$] \textbf{{Extensions of \gls{S-RAPS} to external schedulers}}: 
We implement extensions to integrate external scheduling simulators

\item[$C_4$] \textbf{{Extension of \gls{S-RAPS} to evaluate incentive structures for scheduling}:} case-study on exploring incentive structures and impact of imposed reward
metrics on workloads and systems.

\item[$C_5$] \textbf{{Development of a \gls{ML} guided scheduling for \gls{S-RAPS}}:} Prototype and proof of concept for \gls{ML} guided scheduling given the dataloaders, datasets and scheduler enabled \gls{S-RAPS}.

\end{description}

\subsection{Computational Artifacts}

\gls{S-RAPS} is an extension of \gls{exadigit}/\gls{RAPS} which was previously released as listed in $A_1$.
All contributions are developed as a branch and evolution of \gls{RAPS}.
\begin{description}
\item[$A_1$] https://code.ornl.gov/exadigit/raps
\item[~]\;(https://doi.org/10.11578/dc.20240627.4)
\item[$A_2$] https://code.ornl.gov/exadigit/raps/-/tree/S-RAPS
\item[$A_3$] https://code.ornl.gov/exadigit/raps/-/tree/fastsim-integration
\item[$A_4$] https://code.ornl.gov/exadigit/raps/-/tree/arunavo
\end{description}

\noindent
The artifacts $A_2$, $A_3$, $A_4$ will eventually be merged into $A_1$. 
As $A_1$ is the basis for all of this, and none of the artifacts and contributions would be possible with it alone.
However as it is not a direct contribution of this work, it will be omitted from the further descriptions, but is listed for completeness.


\bigskip
\begin{center}
\begin{tabular}{rll}
\toprule
Artifact ID  &  Contributions &  Related \\
             &  Supported     &  Paper Elements \\
\midrule
$A_2$   & $C_1$, $C_2$ & Figures 4-6 \\
        & $C_4$ & Figure 8 \\
        & $C_3$ & (Sec. 4.2.1) \\
 \midrule
$A_3$   &  $C_3$ & Figure 7 \\
        &        & \\
\midrule
$A_4$   &  $C_5$ & Figure 10 \\
\bottomrule
\end{tabular}
\end{center}

\section{Artifact Identification}
\newartifact
Base branch of the repository. 
\artrel
As discussed, this is no new artifact, but the artifact that the artifacts $A_2$, $A_3$, $A_4$ will be merged into. This is present for future reference.
(Therefore Expected Results, Reproduction Time, Setup, Execution and Analysis are skipped for, and referred to in the other artifacts.)

\newartifact
Implementation of \textbf{\Gls{S-RAPS}}. This artifact introduces scheduling capabilities to \gls{exadigit}, specifically to \gls{exadigit}'s \gls{RAPS}. 
The artifact $A_2$ implements the design as specified in Section~3.
The artifact adds the new engine loop, adds dataloaders and implements the build-in and interface to the external scheduler ScheduleFlow.

\artrel
Main branch of the work presented in the paper, enabling $C_1$, $C_2$, $C_3$, $C_4$, and forming the major rework, enabling $A_3$ and $A_4$ with their Contribution to $C_3$ and $C_5$.


\artexp
The artifact lets you run the \gls{DCDT} simulations as implemented in \gls{S-RAPS}.
This allows to reproduce [Figure~4][Figure~5] as well as (with other datasets than in the paper) [Figure~6], and [Figure~8].
\begin{itemize}
    \item Running \gls{S-RAPS} with synthetic data
    \item Running \gls{S-RAPS} with data from
    \begin{itemize}
        \item Marconi100 (PM100 Dataset) [Figure 4]
        \item Adastra (Cirou's Dataset) 
        \item Fugaku (F-Data) [Figure 5]
        \item Lassen (LAST Dataset)
        \item Frontier [Figure 6]
    \end{itemize}
    \item Running scheduling policies
        \textit{replay},
        \textit{\gls{fcfs}},
        \textit{priority}, and
        backfill policies
        \textit{first-fit}, and 
        \textit{\gls{easy}}.  [Figures 4-6]
    \item Loading and saving accounts.json files for and scheduling using priorities by account using
       \textit{acct\_avg\_power},\\
       \textit{acct\_low\_avg\_power},
       \textit{acct\_edp},
       \textit{acct\_fugaku\_pts} policies,
       in combination with backfill policies. [Figure 8.]
\end{itemize}


\arttime

Time Breakdown: Note: Simulation times depend on selected datasets and job queue lenght!!
\begin{itemize}
  \item Setup (clone + submodule): $<$1 min  
  \item Setup: collecting of datasets ($\sim$4GB): $\sim$25 min
  \item Execution by Figure:
    \begin{itemize}
      \item Figure 4: 4 runs x 3 min -- $\sim$12 min
      \item Figure 5: 4 runs x 25 min -- $\sim$ 100 min
      \item Figure 6: 4 runs x 4 min -- $\sim$12 min 
      \item Figure 8: 5 runs x 3 min -- $\sim$15 min;\\
        alternative reproduction with PM100 data:\\
              Figure 8 alt: 5 runs x 3 min -- $\sim$15 min
    \end{itemize}
  \item Post-Processing (plotting): 20 minutes.
\end{itemize}
Total:  $\sim$ 185 minutes 

Note: Estimate includes plot for each Figure or the alternative, even if Figure 6, will not be reproducible. As Figure 4, 5 and 6 are similar, no alternative for Figure 6 is given, even if it includes the additional cooling model. No open cooling model was available at the time of writing, e.g. for PM100. 

For future reference reach out to \url{exadigit.github.io}, and\\
\url{https://code.ornl.gov/exadigit/datacenterCoolingModel}
or \\
\url{https://code.ornl.gov/exadigit/autocsm}

\artin

\artinpart{Hardware} Laptop capable of running python 3.9+

\artinpart{Software} Python 3.9+
\begin{itemize}
\item git (2.39.5)
\item raps (branch: \url{https://code.ornl.gov/exadigit/raps/-/tree/S-RAPS} )
\item Python 3.9+ (Tested until 3.13)
\item pip 25.0.1
\item Python libraries (Install via pip and raps/\lstinline{pyproject.toml}):
    \begin{itemize}
        \item matplotlib (3.7.2)
        \item numpy (1.23.5)
        \item rich (13.6.0)
        \item fmpy(0.3.19)
        \item pandas (2.0.3)
        \item scipy (1.10.1)
        \item pyarrow (15.0.1)
        \item tqdm (4.66.5)
        \item requests (2.32.3)
    \end{itemize}
\end{itemize}

\artinpart{Datasets / Inputs}

\begin{itemize}
    \item Marconi100: \url{https://zenodo.org/records/10127767/files/job_table.parquet}
    \item Adastra: \url{https://zenodo.org/records/14007065/files/AdastaJobsMI250_15days.parquet}
    \item Fugaku: \url{https://zenodo.org/records/11467483/files/24_04.parquet}
    \item Lassen: \url{https://github.com/LLNL/LAST/tree/main/Lassen-Supercomputer-Job-Dataset}
    \item Frontier: Not released.
\end{itemize}
Descriptions for usage of the datasets are part of the dataloaders in \gls{S-RAPS} directory: \lstinline{raps/dataloaders}.

\artinpart{Installation and Deployment}
\begin{itemize}
  \item Load Python 3.9 (if not installed use conda or something better e.g.: \url{https://github.com/maiterth/pyact}
  \item Pip install python dependencies
  \item Download datasets e.g. to {\$HOME/data/<system>}
\end{itemize}


\artcomp
Artefact Execution by Figure / Paper Element:

\begin{itemize}
    \item For Figure 4: 
        the Artifact $A_2$ is executed with the Marconi System, loading its PM100 dataset. The fastforward is set to 4381000 seconds simulation time is set to 61000 seconds.
        Then the simulation is run 4 times. For each of these task, the same settings are run except:
        \begin{itemize}
            \item[$T_1$:] Marconi100 simulation with \texttt{replay} policy (default)%
            \item[$T_2$:] Marconi100 simulation with \texttt{fcfs} policy + no backfill
            \item[$T_3$:] Marconi100 simulation with \texttt{fcfs} policy\\ + \texttt{easy} backfill
            \item[$T_4$:] Marconi100 simulation with \texttt{priority} policy\\ + \texttt{first-fit} backfill.
        \end{itemize}
        Each generated output directory is copied to a common directory named after the policy.
        Finally the figure is plotted:
        \begin{itemize}
            \item[$T_5$:] The plotting script to generate the Figure 4 is run with the directory as input.
        \end{itemize}
    \item [] Task Dependency: [$T_1,T_2,T_3,T_4,] \rightarrow T_5$ \\
    \item For Figure 5.:
        the Artifact $A_2$ is executed with the Adastra System, loading Cirou's dataset fully.
        Then the simulation is run 4 times. For each of these task, the same settings are run except:
        \begin{itemize}
            \item[$T_6$:] Adastra simulation with \texttt{replay} policy (default)
            \item[$T_7$:] Adastra simulation with \texttt{fcfs} policy
            \item[$T_8$:] Adastra simulation with \texttt{fcfs} policy + \texttt{easy} backfill
            \item[$T_9$:] Adastra simulation with \texttt{priority} policy + \texttt{first-fit} backfill
        \end{itemize}
        Each generated output directory is copied to a common directory named after the policy.
        Finally the figure is plotted:
        \begin{itemize}
            \item[$T_{10}$:] The plotting script to generate the Figure 5 is run with the directory as input.
        \end{itemize}
    \item [] Task Dependency: [$T_6,T_7,T_8,T_9,] \rightarrow T_{10}$ \\

    \item For Figure 6.:
        the Artifact $A_2$ was executed with the Frontier System, with a dataset not publicly available. 
        The setup is as before, only that a cooling model is loaded using the \texttt{-c} option.
        The plotting script available in the Repository, but as the data is not available, this is not reproducible directly. 
        For equivalent plots without PUE and Cooling Tower Temperature see \textit{AD} Figure 4 and~5. 
    \item For Figure 8.:
        For Fig. 8 the Artifact $A_2$ was executed with the Frontier System, and is therefore also not reproducible. However we provide an alternative:
    \item Figure 8 alternative:   
        uses the same steps as Figure 8 but with the Marconi dataset. The result may not as easily readable as the result shown in the paper, but the steps are provided to show the capabilities and potential reproduction or transfer for other users.\\        
        The alternative is executed with the Marconi System, loading its PM100 dataset as described for Figure 4.
        The fastforward option is set to 4381000 seconds,  simulation time is set to 61000 seconds.\\ 
        The simulation is run 5 times. For each of these task, the same settings are run except:
        \begin{itemize}
            \item [$T_{11}$] Run Marconi100 simulation with \texttt{replay} policy (default) adding the \code{\lstinline{--accounts}} option  (save or copy to output folder printed on the screen to a known location).
            \item [$T_{12}$] Use the output directory of $T_{11}$ as input for the recorded account behavior. For $T_{13}$ - $T_{16}$.
            \item[$T_{13}$:] Run Marconi100 simulation with \code{\lstinline{acct_avg_power}} policy + \texttt{first-fit} backfill and reload the accounts from $T_{11}$
            \item[$T_{14}$:] Marconi100 simulation with \texttt{\lstinline{acct_low_avg_power}} policy + \texttt{first-fit} backfill and reload the accounts from $T_{11}$
            \item[$T_{15}$:] Run Marconi100 simulation with \texttt{\lstinline{acct_edp}} policy + \texttt{first-fit} backfill and reload the accounts from $T_{11}$
            \item[$T_{16}$:] Run Marconi100 simulation with \texttt{\lstinline{acct_fugaku_pts}} policy + \texttt{first-fit} backfill and reload the accounts from $T_{11}$
        \end{itemize}

        and for each simulation the output directory is copied to a common directory which is next given to the plotting script.
        \begin{itemize}
            \item[$T_{17}$:] The plotting script for Figure 8 is run.
        \end{itemize}
        \item[] Task Dependency: $T_{11} \rightarrow [T_{12},T_{13},T_{14},T_{15},T_{16}] \rightarrow T_{17}$\\
    \item For $C_3$ -- Paragraph on ScheduleFlow:
        For $C_3$ the Artifact $A_2$ is run using the ScheduleFlow scheduling simulator, with option \texttt{\lstinline{--scheduler scheduleflow}}.
        Note: as the simulation takes very long, therefore abort with \texttt{\lstinline{^C}} or set the simulation time to a low value. 
        With default parameters (generating synthetic jobs) and time limit of \texttt{\lstinline{-t 1h}} the simulation usually terminates within $\sim$2 minutes, depending on the number of random jobs generated.
\end{itemize}




\artout
The artifact outputs are the folders containing
\begin{itemize}
    \item Screen output: Each simulation shows an interactive TUI while running the simulation, showing statistics of the run. Addtitionally a line is printed informing of the output directory of files (when specifying the \texttt{\lstinline{-o}} flag.
    \item Files: the files in the output directory are:\\
    (e.g. \texttt{\lstinline{simulation\_results/<7-dig-hex>}})
        \begin{itemize}
            \item \texttt{\lstinline{acounts.json}} (if \texttt{\lstinline{--accounts}} was specified)

            \item \texttt{\lstinline{job_history.csv}}
            \item \texttt{\lstinline{loss_history.parquet}}
            \item \texttt{\lstinline{power_history.parquet}}
            \item \texttt{\lstinline{queue_history.csv}}
            \item \texttt{\lstinline{running_history.csv}}
            \item \texttt{\lstinline{stats.out}}
            \item \texttt{\lstinline{util.parquet}}
        \end{itemize}
        Additional, if a cooling model is present (not generally available) the options \texttt{\lstinline{-c}} combined with \texttt{\lstinline{-o}} generate the following files:
        \begin{itemize}
            \item \texttt{\lstinline{cooling_model.parquet}}
        \end{itemize}
    \item Plots: If the according plotting scripts in the \texttt{\lstinline{scripts/}} folder were run, graphs for utilization, power, etc. as seen in the publication are generated.
\end{itemize}

Specifically:
\begin{itemize}
    \item Figure 4 (exact reproduction possible)
    \item Figure 5 (exact reproduction possible
    \item Figure 6 (Public  reproduction not possible due do non-public data \& cooling model)
    \item Figure 8 (Public reproduction not possible due do non-public data)
    \item Figure 8 alternative
\end{itemize}

\newartifact

Extension of S-RAPS to interface with the FastSim scheduling simulator. This is a branch of the S-RAPS repository with the integration enabled.
The FastSim scheduler is not open at the time of this writing, and parts of this artifcat referring to the operation of the FastSim scheduler are descriptive only. However, the results of the simulation are available as a dataset, and the code used to generate the synthetic dataset used in the simulation and the plot in Figure 7 are also made available.


\artrel
This artifact directly contributes to $C_3$, and produced Figure 7. The integration of FastSim with S-RAPS is a specific instance of the extension of S-RAPS to external schedulers. 


\artexp
This experiment progresses in four stages: synthetic job trace generation, simulation with FastSim, replay with S-RAPS, and plotting the results as a final figure, which result in a synthetic job trace, a rescheduled job trace, an output file from \gls{S-RAPS}, and Figure 7, respectively. These results substantiate $C_3$ by demonstrating the integration of FastSim with \gls{S-RAPS}.



\arttime

Time Breakdown:
\begin{itemize}
    \item Setup (Switch branch and clone): 30 seconds
    \item Execution (Data-set generation): 10 seconds
    \item Execution (FastSim Simulation Time): 16 minutes
    \item Execution (FastSim Output Transformation): 30 seconds
    \item Execution (S-RAPS Replay Time): 15 minutes
    \item Postprocessing (Figure 7): Less than 1 minute
\end{itemize}



\artin

\artinpart{Hardware}
See $A_2$.

\artinpart{Software}
See $A_2$. The software setup follows the same procedure, while only changing the branch to: \url{https://code.ornl.gov/exadigit/raps/-/tree/fastsim-integration}.

\artinpart{Datasets / Inputs}
As FastSim is not open, the entire process cannot be run from start to finish. In its place, datasets have been provided for each milestone. Each of these datasets, and supplemental Python scripts associated with this artifact, are found in the RAPS repository branch associated with FastSim integration ($A_3$), in the directory \texttt{raps/schedulers/fastsim\_artifact}. The available datasets are:

\begin{itemize}
    \item \texttt{sacct\_jobs.csv}: the synthetic job trace used by FastSim
    \item \texttt{frontier\_simulation\_results.csv}: the simulated job trace as scheduled by FastSim
    \item \texttt{jobs.parquet}: the first part of the transformed FastSim output data required by the RAPS Frontier dataloader 
    \item \texttt{jobsprofile.parquet}: the second part of the transformed FastSim output data required by the RAPS Frontier dataloader
    \item \texttt{raps\_results.csv}: the output of the RAPS replay of the FastSim schedule simulation
\end{itemize}

\artinpart{Installation and Deployment}

\begin{itemize}
    \item Switch to fastsim-integration branch.
    \item Install dependencies (same as $A_2$)
\end{itemize}

\artcomp

\begin{itemize}
    \item{$T_1$: } Create synthetic job trace with: \\ 
    \hspace*{1.9em} \texttt{synthetic\_jobs.py}
    \item{$T_2$: } Import synthetic job trace and simulate scheduling with FastSim in \texttt{plugin-mode}
    \item{$T_3$: } Transform FastSim output to format required by RAPS Frontier dataloader with: \\ 
    \hspace*{1.9em} \texttt{transform\_data.py}
    \item{$T_4$: } Run RAPS in replay mode with the following command: \\
    \texttt{python main.py -f ./raps/jobs.parquet \\\hspace*{1.9em} ./raps/jobsprofile.parquet}
    \item{$T_5$: } Plot output results from RAPS replay with: \\ \hspace*{1.9em} \texttt{plot\_results.py}
\end{itemize}

\artout

The final RAPS simulation produces an output directory with the files and time-series (using the \texttt{\lstinline{-o}} flag) which can be plotted as seen in Figure 7.


\newartifact
Implementation of ML component of \textbf{\gls{S-RAPS}}. This artifact introduces ML-guided scheduling capabilities to \gls{exadigit}, specifically to \gls{exadigit}'s \gls{RAPS}. 
The artifact $A_4$ implements the design as specified in Section~4.4.
The artifact adds the new ML component, including novel predictor models and classifiers, dataloaders, and implements the scoring component to the external scheduler ScheduleFlow.


\artrel
This component relates to the ML-based scheduling policy evaluation use case presented in the paper, enabling $C_5$.
\artexp

With ML-guided scheduling, the artifact lets you run the \gls{DCDT} simulations as implemented in \gls{S-RAPS}.
This allows for the reproduction of [Figure~10]. Depending on the scoring component, the ML-guided scheduling is expected to perform better than the other existing scheduling algorithms in one way or another. These experiments show how \gls{S-RAPS} accommodates exploration of various ML-guided scheduling policies by the community. 
\begin{itemize}
    \item Running \gls{S-RAPS} with data from
    \begin{itemize}
        \item Fugaku (F-Data) [Figure 10]    
    \end{itemize}
    \item Running scheduling policies
        \textit{\gls{fcfs}},
        \textit{priority}, 
        \textit{sjf},
        \textit{ljf},
        \textit{ML}, with 
        backfill policy
        \textit{firstfit}.  [Figure 10]
    
\end{itemize}



\arttime

\begin{itemize}
  \item Setup (clone): $<$1 min  
  \item Setup: collecting Fugaku datasets -- ($\sim$2GB): $\sim$2 min
  
    Files: \begin{itemize}
      \item 21\_03.parquet
      \item 21\_04.parquet
      \item 21\_05.parquet
      \item 21\_06.parquet
      \item 21\_07.parquet
      \item 21\_08.parquet
  \end{itemize}

  \item Execution of Figure 10:
    \begin{itemize}
        \item Data split and Preprocessing -- $\sim$4 min
        \item Training -- $\sim$27 min
        \item Inference -- $\sim$1 min
        \item \gls{S-RAPS} -- 5 runs x 40 min -- $\sim$200 min
    \end{itemize}
  \item Post-Processing (plotting): $\sim$ 2 minutes.
\end{itemize}
Total: $\sim$ 230 minutes



\artin

\artinpart{Hardware} Laptop capable of running python 3.9+

\artinpart{Software} Python 3.9+
\begin{itemize}
\item git (2.39.5)
\item raps (branch: \url{https://code.ornl.gov/exadigit/raps/-/tree/ml-guided-scheduling} )
\item Python 3.9+
\item pip 25.0.1
\item Python libraries (Install via pip and raps/\lstinline{pyproject.toml}):
    \begin{itemize}
        \item joblib (1.4.2)
        \item scikit-learn (1.6.1)
        \item pyyaml (6.0.2)
        \item jsonlines (4.0.0)
        \item plotly (6.0.1)
        \item kaleido (0.2.1)
        \item matplotlib (3.7.2)
        \item numpy (1.23.5)
        \item rich (13.6.0)
        \item fmpy(0.3.19)
        \item pandas (2.0.3)
        \item scipy (1.10.1)
        \item pyarrow (15.0.1)
        \item tqdm (4.66.5)
        \item requests (2.32.3)
    \end{itemize}
\end{itemize}

\artinpart{Datasets / Inputs}

Fugaku:
\begin{itemize}
    \item \url{https://zenodo.org/records/11467483/files/21_03.parquet}
    \item \url{https://zenodo.org/records/11467483/files/21_04.parquet}
    \item \url{https://zenodo.org/records/11467483/files/21_05.parquet}
    \item \url{https://zenodo.org/records/11467483/files/21_06.parquet}
    \item \url{https://zenodo.org/records/11467483/files/21_07.parquet}
    \item \url{https://zenodo.org/records/11467483/files/21_08.parquet}
\end{itemize}

The configuration file defining clustering methods, classifiers, and prediction parameters for various use cases is located within the app-fingerprinting module, under the \gls{S-RAPS} repository at \lstinline{app-fingerprinting/config}.

\artinpart{Installation and Deployment}
\begin{itemize}
  \item Load Python 3.9 
  \item \begin{verbatim}
> cd app-fingerprinting
> python3.9 -m venv venv
> source venv/bin/activate 
> pip install -r requirements.txt
> cd ..
\end{verbatim}
  \item Download Fugaku datasets to {./app-fingerprinting/data/<system>}
\end{itemize}

\artcomp
Artifact Execution of Figure 10:

The Artifact $A_4$ is executed with the F-Data dataset.

\subsubsection{Training [$T_1$:]}
\begin{itemize}
    \item [$T_{1.1}$:] Split the F-data into train and test datasets.
    \item[$T_{1.2}$:] Pre-process the train and test datasets
    \item[$T_{1.3}$:] Cluster the train dataset into 5 clusters
    \item[$T_{1.4}$:] Train the classifier with pre-submission features and cluster labels
    \item[$T_{1.5}$:] Train regression model with pre-submission features and target features(resource usage, duration)
\end{itemize}

Task dependency: $T_{1.1} \rightarrow T_{1.2} \rightarrow T_{1.3} \rightarrow T_{1.4}  \rightarrow T_{1.5}$ \\

\subsubsection{Inference [$T_2$:]}
\begin{itemize}
    \item [$T_{2.1}$:] Load Preprocessed Test dataset
    \item [$T_{2.2}$:] Classify them into clusters 
    \item [$T_{2.3}$:] Use the corresponding model to predict resource usage and duration of the job in the test dataset
    \item [$T_{2.4}$:] Rank the jobs based on predicted resource usage, duration, and pre-submitted features of a job
\end{itemize}

Task dependency: $T_{2.1} \rightarrow T_{2.2} \rightarrow T_{2.3} \rightarrow T_{2.4} $ \\

\subsubsection{\gls{S-RAPS} [$T3$:]}

\begin{itemize}

\item [$T_{3.1}$:] Load test dataset with ranking scores. 

\end{itemize}

The fast-forward is set to 35 days, and the simulation time is set to 7 days. The simulation was run from 2021-04-05 to 2021-04-11.
Then the simulation is repeated 5 times. For each of these tasks, the same settings are run except for the policy:
    \begin{itemize}
        \item[$T_{3.2}$:] F-Data simulation with \texttt{fcfs} policy\\ +  \texttt{first-fit} backfill.
        \item[$T_{3.3}$:] Fugaku simulation with \texttt{sjf} policy\\ +  \texttt{first-fit} backfill.
        \item[$T_{3.4}$:] F-Data simulation with \texttt{priority} policy\\ + \texttt{first-fit} backfill.
        \item[$T_{3.5}$:] F-Data simulation with \texttt{ljf} policy\\ + \texttt{first-fit} backfill.
        \item[$T_{3.6}$:] F-Data simulation with \texttt{ML-guided} policy\\ + \texttt{first-fit} backfill. This policy uses ranks calculated for each job in the inference stage.
    \end{itemize}
    
Each generated output directory can be renamed to a directory named after the policy.

All the above output directories are copied into a common directory say latest\_run and this directory will be input to plotting scripts.
    \begin{itemize}
        \item[$T_{3.7}$:] The plotting script to generate the Figure 10 is run with the latest\_run directory as input.
    \end{itemize}

Task dependency: $T_{3.1} \rightarrow [T_{3.2}, T_{3.3},  T_{3.4}, T_{3.5}, T_{3.6}] \rightarrow T_{3.7} $ \\

Overall Task Dependency: $T_1 \rightarrow T_2 \rightarrow T_3$ \\
    
\artout
We have 3 phases. Each phase will have different output folders. In \texttt{app-fingerprinting/}, the output is stored in \texttt{models/} and \texttt{results/} directory.

\subsubsection{Training}
    \begin{itemize}
        \item [$T_{1.1}$:] The raw data set is split into train and test dataset. The outputs are stored in the following files.

        \begin{itemize}
            \item app-fingerprinting/results/<system>/\\
        $\hookrightarrow$<system>\_train.parquet
            \item app-fingerprinting/results/<system>/\\
        $\hookrightarrow$<system>\_test.parquet
        \end{itemize}
        
        \item[$T_{1.2}$:] The test and train datasets are preprocessed separately and stored in the following files.

        \begin{itemize}
            \item app-fingerprinting/results/<system>/\\
            $\hookrightarrow$<system>\_train\_preprocessed.parquet
            \item app-fingerprinting/results/<system>/\\
            $\hookrightarrow$<system>\_test\_preprocessed.parquet
        \end{itemize}
        
        \item[$T_{1.3}$:] The pre-processed train dataset is clusterd into 5 clusters. The cluster labels and their corresponding models are stored in a pkl file.  
        \begin{itemize}
            \item app-fingerprinting/models/<system>/\\
        $\hookrightarrow$clusterer/<cluster\_model>.pkl
        \end{itemize}
        The pkl file contains 
        \begin{itemize}
            \item cluster labels
            \item cluster model
            \item cluster algorithm name 
        \end{itemize}
        
        \item[$T_{1.4}$:] The classifier is trained and stored in the a pkl file as well for future use during inference.
        \begin{itemize}
            \item app-fingerprinting/models/<system>/\\
        $\hookrightarrow$cluster\_classifiers/<classifier>.pkl
            \end{itemize}
        This pkl file contains 
        \begin{itemize}
            \item Trained classifier model that can classify new jobs during inference \end{itemize}
        
        \item[$T_{1.5}$:] A prediction model for each cluster is trained and saved as
        \begin{itemize}
        \item app-fingerprinting/models/<system>/prediction\_models/<model>/\\
        $\hookrightarrow$<cluster\_number>.pkl
        \end{itemize}
        
        Each of the pkl files contains 
        \begin{itemize}
            \item Trained regressor model
            \item Cluster label ID
        \end{itemize}
    \end{itemize}
    
\subsubsection{Inference}

The output of this phase is a single parquet file which is a raw test dataset along with ranking scores respectively.

The output file is stored in 
\begin{itemize}
    \item app-fingerprinting/results/<system>/inference\_results.parquet 
\end{itemize}

\subsubsection{\gls{S-RAPS}}

\begin{itemize}
    \item Screen output: Each simulation shows an interactive GUI while running the simulation, showing statistics of the run. Additionally a line is printed informing of the output directory of files (when specifying the \texttt{\lstinline{-o}} flag.
    \item Files: the files in the output directory are:\\
    (e.g. \texttt{\lstinline{simulation\_results/<7-dig-hex>}})
        \begin{itemize}
            \item \texttt{\lstinline{acounts.json}} (if \texttt{\lstinline{--accounts}} was specified)

            \item \texttt{\lstinline{job_history.csv}}
            \item \texttt{\lstinline{loss_history.parquet}}
            \item \texttt{\lstinline{power_history.parquet}}
            \item \texttt{\lstinline{queue_history.csv}}
            \item \texttt{\lstinline{running_history.csv}}
            \item \texttt{\lstinline{stats.out}}
            \item \texttt{\lstinline{util.parquet}}
        \end{itemize}
    \item Plots: 
        \begin{itemize}
            \item
            For power plot, the \texttt{\lstinline{power\_history.parquet}} file of all policies is given as input to \texttt{plot\_power.py} script
            \item 
            For metrict radar chart, the \texttt{\lstinline{stats.out}} file of all policies is given as input to \texttt{\lstinline{plot\_stats.py}} script
            \end{itemize}
            
        These scripts are present in the folder \texttt{\lstinline{scripts/}} of app-fingerprinting sub module. 
        
\end{itemize}

\clearpage
\appendixAE
\arteval{1} Omitted acc. to AD.

\arteval{2}
\artin
Instructions for $A_2$
\begin{itemize}
    \item \texttt{\lstinline{> git clone https://code.ornl.gov\\}}\\
    \texttt{\lstinline{/exadigit/raps/-/tree/S-RAPS}}
    \item \texttt{\lstinline{> cd raps}}
    \item Install and load python3.9\\
        if not installed use conda or something better\\
        e.g.: \url{https://github.com/maiterth/pyact.git}
    \item Make sure pip in installed and install dependencies:
\begin{verbatim}
> pip install -e .
\end{verbatim}
    \item Download Datasets:
    \begin{itemize}
        \item Create a directory for the data e.g.
\begin{verbatim}
> mkdir ~/data && cd ~/data
\end{verbatim}
        \item Marconi100:
\begin{verbatim}
> mkdir marconi100 && cd marconi100
> wget https://zenodo.org/records/\
10127767/files/job_table.parquet
\end{verbatim}
        \item Adastra: 
\begin{verbatim}
> mkdir marconi100 && cd marconi100
> wget https://zenodo.org/records/\
14007065/files/AdastaJobsMI250_15days.parquet
\end{verbatim}
        \item Fugaku:
\begin{verbatim}
> mkdir fugaku && cd fugaku
> wget https://zenodo.org/records/11467483\
/files/24_04.parquet
\end{verbatim}
        \item Lassen:
\begin{verbatim}
> mkdir lassen && cd lassen
> git clone https://github.com/LLNL/LAST/ 
> cd LAST
> git lfs pull
\end{verbatim}
    \end{itemize}
    \item You're ready to go!
\end{itemize}


\artcomp
Computation according to paper element:
\begin{itemize}
  \item Figure 4:
  \item Figure 5:
  \item Figure 6:
  \item Figure 8:
  \item Sec 4.2.1 /$C_3$:
\end{itemize}
In the following we show how to execute the commands to generate the figures:

Assuming your data is in \texttt{\lstinline{\$HOME}}:
\begin{itemize}
  \item For Figure 4:
  \begin{itemize}
    \item Run\\{
    \texttt{\lstinline{> python main.py -f \\}}\\
    \texttt{\lstinline{\~/data/marconi100/job\_table.parquet \\}}\\
    \texttt{\lstinline{--system marconi100 -o -ff 4381000 \\}}\\
    \texttt{\lstinline{-t 61000 --scheduler default --policy replay}}
    }
    \item Run\\{
    \texttt{\lstinline{> python main.py -f \\}}\\
    \texttt{\lstinline{\~/data/marconi100/job\_table.parquet \\}}\\
    \texttt{\lstinline{--system marconi100 -o -ff 4381000 \\}}\\
    \texttt{\lstinline{-t 61000 --scheduler default --policy fcfs}}
    }
    \item Run\\{
    \texttt{\lstinline{> python main.py -f \\}}\\
    \texttt{\lstinline{\~/data/marconi100/job\_table.parquet \\}}\\
    \texttt{\lstinline{--system marconi100 -o -ff 4381000 \\}}\\
    \texttt{\lstinline{-t 61000 --scheduler default --policy fcfs \\}}\\
    \texttt{\lstinline{--backfill easy}}
    }
    \item Run\\{
    \texttt{\lstinline{> python main.py -f \\}}\\
    \texttt{\lstinline{\~/data/marconi100/job\_table.parquet \\}}\\
    \texttt{\lstinline{--system marconi100 -o -ff 4381000 \\}}\\
    \texttt{\lstinline{-t 61000 --scheduler default --policy \\}}\\
    \texttt{\lstinline{priority --backfill first-fit}}
    }
    \item For each of the commands above copy the generated output directories to a common directory \texttt{\lstinline{<dir>}} which you each named after its policy (\texttt{\lstinline{replay}},
    \texttt{\lstinline{fcfs-nobf}},
    \texttt{\lstinline{fcfs-easy}},
    \texttt{\lstinline{priority-ffbf}} )
    \item Run:\\
    \texttt{\lstinline{python scripts/plots/2in1-pm100day50.py <dir>}}
  \end{itemize}
  \item For Figure 5:
  \begin{itemize}
    \item Run\\{
    \texttt{\lstinline{python main.py --system adastraMI250 -f \\}}\\
    \texttt{\lstinline{\~/data/adastra/AdastaJobsMI250\_15days.parquet \\}}\\
    \texttt{\lstinline{-o --scheduler default --policy replay}}
    }
    \item Run\\{ 
    \texttt{\lstinline{python main.py --system adastraMI250 -f \\}}\\
    \texttt{\lstinline{\~/data/adastra/AdastaJobsMI250\_15days.parquet \\}}\\
    \texttt{\lstinline{-o --scheduler default --policy fcfs}}
    }
    \item Run\\{
    \texttt{\lstinline{python main.py --system adastraMI250 -f \\}}\\
    \texttt{\lstinline{\~/data/adastra/AdastaJobsMI250\_15days.parquet \\}}\\
    \texttt{\lstinline{-o --scheduler default --policy fcfs \\}}\\
    \texttt{\lstinline{--backfill easy}}
    }
    \item Run\\{
    \texttt{\lstinline{python main.py --system adastraMI250 -f \\}}\\
    \texttt{\lstinline{\~/data/adastra/AdastaJobsMI250\_15days.parquet \\}}\\
    \texttt{\lstinline{-o --scheduler default --policy easy \\}}\\
    \texttt{\lstinline{--backfill first-fit}}
    }
    \item For each of the commands above copy the generated output directories to a common directory \texttt{\lstinline{<dir>}} which you each named after its policy (\texttt{\lstinline{replay}},
    \texttt{\lstinline{fcfs-nobf}},
    \texttt{\lstinline{fcfs-easy}},
    \texttt{\lstinline{priority-ffbf}})
    \item Run: \\
    \texttt{\lstinline{python scripts/plots/2in1-adastra.py <dir>}}
  \end{itemize}
  
  \item For Figure 6:
    This is not directly reproducible as the data and cooling model used is not publicly available. With available cooling model, the procedure to generate is similar to the ones of Figure 4 and 6.
  \begin{itemize}
      \item To Run add the \texttt{\lstinline{-c}} option if a cooling model is available. (e.g. \texttt{\lstinline{python main.py -c ...}}).
      \item Use the same four runs as shown in the above artifact evaluations for Figures 4 and 6. 
      \item The plotting-script is available in\\
      \texttt{\lstinline{./scripts/plots/4in1-frontier-wC.py}}. 
  \end{itemize}
  \item For Figure 8:
   Similar to Figure 6, this  is not directly reproducible as the data used is not publicly available. An equivalent reproduction is explained in Figure 8 alternative.
  \item For Figure 8 alternative:
    \begin{itemize}
    \item Run\\{
        \texttt{\lstinline{> python main.py -f \\}}\\
        \texttt{\lstinline{\~/data/marconi100/job\_table.parquet \\}}\\
        \texttt{\lstinline{--system marconi100 -o -ff 4381000 \\}}\\
        \texttt{\lstinline{-t 61000 --scheduler default --policy replay \\}}\\
        \texttt{\lstinline{--accounts}}
    }
    \item Save the output directory to a common directory for all the outputs and name it \texttt{\lstinline{<replay>}} (placeholder). This directory is now used to read in the accounts statistics for the subsequent runs.
    \item Run\\{
        \texttt{\lstinline{> python main.py -f \\}}\\
        \texttt{\lstinline{\~/data/marconi100/job\_table.parquet \\}}\\
        \texttt{\lstinline{--system marconi100 -o -ff 4381000 \\}}\\
        \texttt{\lstinline{-t 61000 --scheduler experimental \\}}\\
        \texttt{\lstinline{--scheduler experimental \\}}\\
        \texttt{\lstinline{--policy acct\_avg\_power \\}}\\
        \texttt{\lstinline{--backfill firstfit --accounts \\}}\\
        \texttt{\lstinline{--accounts-json <replay>/accounts.json}}
    }
    \item Run\\{
        \texttt{\lstinline{> python main.py -f \\}}\\
        \texttt{\lstinline{\~/data/marconi100/job\_table.parquet \\}}\\
        \texttt{\lstinline{--system marconi100 -o -ff 4381000 \\}}\\
        \texttt{\lstinline{-t 61000 --scheduler experimental \\}}\\
        \texttt{\lstinline{--scheduler experimental \\}}\\
        \texttt{\lstinline{--policy acct\_low\_avg\_power \\}}\\
        \texttt{\lstinline{--backfill firstfit --accounts \\}}\\
        \texttt{\lstinline{--accounts-json <replay>/accounts.json}}
    }
    \item Run\\{
        \texttt{\lstinline{> python main.py -f \\}}\\
        \texttt{\lstinline{\~/data/marconi100/job\_table.parquet \\}}\\
        \texttt{\lstinline{--system marconi100 -o -ff 4381000 \\}}\\
        \texttt{\lstinline{-t 61000 --scheduler experimental \\}}\\
        \texttt{\lstinline{--scheduler experimental \\}}\\
        \texttt{\lstinline{--policy acct\_edp \\}}\\
        \texttt{\lstinline{--backfill firstfit --accounts \\}}\\
        \texttt{\lstinline{--accounts-json <replay>/accounts.json}}
    }
    \item Run\\{
        \texttt{\lstinline{> python main.py -f \\}}\\
        \texttt{\lstinline{\~/data/marconi100/job\_table.parquet \\}}\\
        \texttt{\lstinline{--system marconi100 -o -ff 4381000 \\}}\\
        \texttt{\lstinline{-t 61000 --scheduler experimental \\}}\\
        \texttt{\lstinline{--scheduler experimental \\}}\\
        \texttt{\lstinline{--policy acct\_fugaku\_pts \\}}\\
        \texttt{\lstinline{--backfill firstfit --accounts \\}}\\
        \texttt{\lstinline{--accounts-json <replay>/accounts.json}}
    }  
    \item For each of the commands above copy the generated output directories to a common directory \texttt{\lstinline{<dir>}} which you each named after its policy (\texttt{\lstinline{replay}},\\
    \texttt{\lstinline{acct\_avg\_power-ffbf}},
    \texttt{\lstinline{acct\_low\_avg\_power-ffbf}},
    \texttt{\lstinline{acct\_edp-ffbf}},
    \texttt{\lstinline{acct\_fugaku\_pts-ffbf}})
    \item Run: \\
    \texttt{\lstinline{python scripts/plots/fgk\_frontier.py <dir>}}
    \end{itemize}

  \item For Sec 4.2.1 /$C_3$ Run:
    \begin{verbatim}
> python main.py -t 1h --scheduler scheduleflow
    \end{verbatim}
    \vspace*{-0.2cm}
    We discovered that scheduleflow may schedule even if nodes are unavailable, which we report as error. This may be a underhandled corner case within scheduleflow, which we check and throw. If this occurs, retry the run, as this does not always occur. Why this discrepancy occurs has yet to be fully debugged.
\end{itemize}

\artout
Each simulation generates an output directory placed in\linebreak
\texttt{\lstinline{simulation_results/}}, as discussed in the execution.
For each generated output directory, the plotting scripts were provided in the analysis step.
We generate a total of 4 runs and one plot for Figure 4
4 simulation runs and one plot for Figure 5
and 5 simulations runs and one plot for the alternative of Figure 8 in this Artifact Evaluation.
This shows the functionality of the simulator and allow us to analyze rescheduling and power simulation of the public datasets as discussed in the paper, supporting contributions $C_1$, $C_2$, $C_3$ and $C_4$.

\arteval{3}
\artin

\begin{itemize}
    \item \texttt{\lstinline{> git clone https://code.ornl.gov\\}}\\
    \texttt{\lstinline{/exadigit/raps/-/tree/fastsim-integration}}
    \item \texttt{\lstinline{> cd raps}}
    \item Install and load Python 3.9
    \item Make sure pip is installed and install dependencies:
\begin{verbatim}
> pip install -e .
\end{verbatim}
\end{itemize}

\artcomp
FastSim is not yet open, so this artifact execution must begin with the provided results of the FastSim scheduling simulation.
\begin{itemize}
 \item Run the RAPS simulation with FastSim output
\begin{verbatim}
> python main.py -f ./raps/jobs.parquet 
  ./raps/jobsprofile.parquet -o --validate
\end{verbatim}
    \item Move the output file (\texttt{job\_history.csv}) from the created directory (see the note at the end of the RAPS simulation) to: \\
    \texttt{raps/raps/schedulers/fastsim\_artifact}
    \item Alternatively, run \texttt{plot\_results.py} directly with \\ \texttt{raps\_results.csv} file.
\end{itemize}

\artout\begin{itemize}
 \item Plot the final results.
\begin{verbatim}
> cd raps/schedulers/fastsim_artifact
> python plot_results.py
\end{verbatim}
\end{itemize}

The power usage demonstrated through this experiment may not be representative of actual Frontier power usage due to differences in a historic job trace and the synthetic job trace created for this artifact. However, the resulting plot mirrors the general power characteristics shown in Figure 7, which demonstrates the integration of external schedulers with S-RAPS ($C_3$).


\arteval{4}
\artin
Instructions for executing the ML-based policy scheduling experiments: 
\begin{itemize}
    \item \texttt{\lstinline{> git clone --recurse-submodules https://code.ornl.gov\\}}\\
    \texttt{\lstinline{/exadigit/raps/-/tree/ml-guided-scheduling}}
    \item Make sure Python3.9 and pip are installed. Then do the following:
\begin{verbatim}
> python3.9 -m venv app-fingerprinting/venv
> source app-fingerprinting/venv/bin/activate  
> pip install -r app-fingerprinting/requirements.txt
\end{verbatim}
    \item Download Datasets:
\begin{verbatim}
> sh app-fingerprinting/scripts/download_fugaku.sh
\end{verbatim}
    \item You're ready to go!
\end{itemize}


\artcomp

Fugaku:
\begin{itemize}
    \item Split and preprocess

\begin{verbatim}
> python3.9 app-fingerprinting/driver.py fugaku split
\end{verbatim}
    \item Train pipeline
\begin{verbatim}
> python3.9 app-fingerprinting/driver.py fugaku train
\end{verbatim}
    \item Inference pipeline
\begin{verbatim}
> python3.9 app-fingerprinting/driver.py fugaku test
\end{verbatim}
    \item For reproducing Figure 10, run \gls{S-RAPS} with 5 different policies
\begin{verbatim}
> python3.9 raps/main.py\
 -f app-fingerprinting/results/\
 fugaku/inference_reselts.parquet \ 
  --system fugaku \
  -ff 35d -t 7d -o \
  --policy fcfs --backfill first-fit
\end{verbatim}

\begin{verbatim}
> python3.9 raps/main.py\
 -f app-fingerprinting/results/\
 fugaku/inference_reselts.parquet \ 
  --system fugaku \
  -ff 35d -t 7d -o \
  --policy sjf --backfill first-fit
\end{verbatim}

\begin{verbatim}
> python3.9 raps/main.py\
 -f app-fingerprinting/results/\
 fugaku/inference_reselts.parquet \ 
  --system fugaku \
  -ff 35d -t 7d -o \
  --policy priority --backfill first-fit
\end{verbatim}

\begin{verbatim}
> python3.9 raps/main.py\
 -f app-fingerprinting/results/\
 fugaku/inference_reselts.parquet \ 
  --system fugaku \
  -ff 35d -t 7d -o \
  --policy ljf --backfill first-fit
\end{verbatim}

\begin{verbatim}
> python3.9 raps/main.py\
 -f app-fingerprinting/results/\
 fugaku/inference_reselts.parquet \ 
  --system fugaku \
  -ff 35d -t 7d -o \
  --policy ml --backfill first-fit
\end{verbatim}

\item Copy the generated output folders from the 5 runs into a directory for post-processing.
\begin{verbatim}
> mkdir latest_run
> ls -dt simulation_results/*/ \
| head -n 5 | xargs -I {} cp -r {} \
latest_run/
\end{verbatim}

\artout

\item Then, for reproducing Figure 10a, do the following: 
\begin{verbatim}
> python3.9 app-fingerprinting/\
scripts/plot_power.py latest_run
\end{verbatim}

\item Plot stats
\begin{verbatim}
> python3.9 app-fingerprinting/\
scripts/plot_stats.py latest_run
\end{verbatim}

\end{itemize}







\end{document}